\documentclass{pasj00}
\color{black}
\tenpoint

\SetRunningHead{K. Sato et al.}{
Suzaku observations of AWM~7}

\title{
Suzaku Observations of AWM~7 Cluster of Galaxies: \\
Temperature, Abundance and Bulk Motions}
\author{
 Kosuke \textsc{Sato},\altaffilmark{1,2}
 Kyoko \textsc{Matsushita},\altaffilmark{1}
 Yoshitaka \textsc{Ishisaki},\altaffilmark{2} 
 Noriko \textsc{Y.~Yamasaki},\altaffilmark{3}\\
 Manabu \textsc{Ishida},\altaffilmark{3} 
 Shin \textsc{Sasaki},\altaffilmark{2}
and Takaya \textsc{Ohashi},\altaffilmark{2} 
}
\altaffiltext{1}{
Department of Physics, Tokyo University of Science,
1-3 Kagurazaka, Shinjuku-ku, Tokyo 162-8601}
\email{ksato@rs.kagu.tus.ac.jp}
\altaffiltext{2}{
Department of Physics, Tokyo Metropolitan University,
 1-1 Minami-Osawa, Hachioji, Tokyo 192-0397}
\altaffiltext{3}{
Institute of Space and Astronautical Science (ISAS),
Japan Aerospace Exploration Agency, \\
3-1-1 Yoshinodai, Sagamihara, Kanagawa 229-8510}
\Received{---}
\Accepted{---}
\Published{---}

\begin{document}
\maketitle

\begin{abstract}
We carried out 3 observations of the cluster of galaxies AWM~7, for
the central region and $20'$-east and $20'$-west offset regions, with
Suzaku.  Temperature and abundance profiles are measured out to
$27'\simeq 570\; h_{70}^{-1}$~kpc, which corresponded to $\sim 0.35\;
r_{180}$.  The temperature of the intra-cluster medium (ICM) slightly
decreases from 3.8~keV at the center to 3.4~keV in $\sim0.35~r_{180}$
region, indicating a flatter profile than those in other nearby
clusters.  Abundance ratio of Si to Fe is almost constant in our
observation, while Mg to Fe ratio increases with radius from the
cluster center. O to Fe ratio in the west region shows increase with
radius, while that in the east region is almost flat, though the
errors are relatively large.  These features suggest that the
enrichment process is significantly different between products of type
II supernovae (O and Mg) and those by type Ia supernovae (Si and Fe).
We also examined positional shift of the central energy of He-like
Fe-K$\alpha$ line, in search of possible rotation of the ICM\@. The
90\% upper limit for the line-of-sight velocity difference was derived
to be $\Delta v \lesssim 2000$ km s$^{-1}$, suggesting that the
ellipticity of AWM~7 is rather caused by a recent directional infall of
the gas along the large-scale filament.
\end{abstract}

\section{Introduction}
Mass profile of the cluster, which is a useful parameter in
constraining the cosmology, is determined through X-ray measurements of
the temperature and density structure of the ICM under the assumption
of hydrostatic equilibrium of the ICM\@.  Recent observations showed
spatially resolved spectra of clusters with good spatial and spectral
resolution, and have revealed the temperature and abundance profiles
for many systems.  \citet{markevitch98} analyzed the spectra of 30
nearby clusters with ASCA and ROSAT, and reported that most of them
showed a similar temperature decline at large radii. However,
\citet{white00} studied a sample of 106 clusters observed with ASCA
and found that 90\% of the temperature profiles were consistent with
being isothermal.  \citet{piffaretti05} showed the temperature
profiles of 13 nearby cooling flow clusters with XMM-Newton, and the
temperature decreased by $\sim30$\% between 0.1 and 0.5 $r_{180}$ using 
the virial radius $r_{180}$ in good agreement with \citet{markevitch98} 
results.  Recent measurementsof 13 nearby relaxed clusters with Chandra 
by \citet{vikhlinin05} also indicated that the temperature reached a peak 
at $r\sim0.15~r_{180}$and then declined to about half of its peak value at
$r\sim0.5~r_{180}$. 

Chandra and XMM-Newton observations also allowed detailed studies of
the metal abundance in the ICM\@.  However, these observations showed
the metal abundance profiles from O to Fe only for the central region
of bright clusters or groups of galaxies dominated by cD galaxies in a
reliable manner
\citep{finoguenov02,fukazawa04,matsushita03,tamura04,matsushita07}.
Abundances of O and Mg in the cluster outer regions are still poorly
determined, because these satellites show relatively high intrinsic
background.  Because of the low and stable background and the good
resolving power for emission lines below 1 keV, Suzaku XIS instrument
\citep{koyama07} can measure not only the precise temperature close to
the virial radius of the cluster, but also abundances of O to Fe to
outer regions \citep{sato07} compared with the past observations.

Clusters are also thought to grow into larger systems through complex
interactions between smaller systems. It is expected that evidence of
merger events can be found as non-Gaussian velocity distributions of
galaxies, temperature and density inhomogeneities in the ICM, and bulk
motions of the ICM\@.  \citet{furusho03} found the inhomogeneous ICM
as blob-like structure in the central region of AWM~7 with Chandra.
\citet{hayakawa06b} also found the inhomogeneities as the high
metallicity blob in the central region of Abell~1060 with Chandra and
XMM-Newton.  Though \citet{dupke06} showed a large velocity gradient
of $2400\pm1000$ km s$^{-1}$ over a spatial scale of 100 kpc in the
Centaurus cluster with Chandra, reconfirming their previous ASCA
measurement \citep{dupke01}, \citet{ota07} gave a negative result of
$\Delta v < 1400$ km s$^{-1}$ based on measurement with Suzaku.

\begin{figure*}
\centerline{
\FigureFile(0.7\textwidth,\textwidth){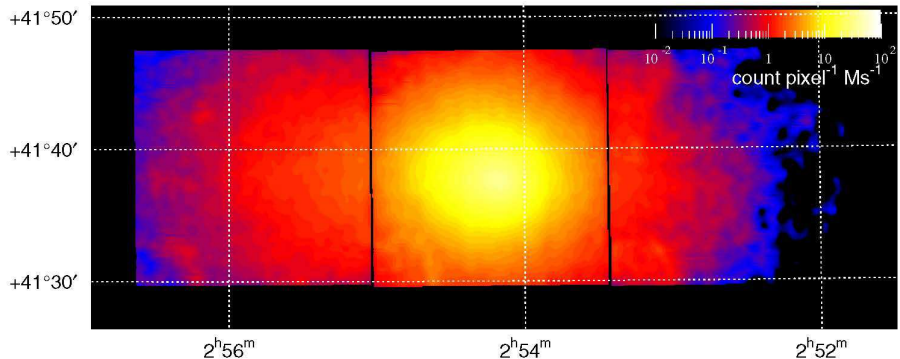}}
\caption{
Combined XIS image of the central and offset observations in the
0.5--7.0 keV energy range. The observed XIS0-3 images are added on the sky
coordinate after removing each calibration source region,
and smoothed with $\sigma=16$ pixel $\simeq 17''$ Gaussian.
Estimated components of extragalactic X-ray background (CXB)
and instrumental background (NXB) are subtracted,
and exposure are corrected, however vignetting is not corrected.
}\label{fig:1}
\end{figure*}

AWM~7 is a nearby cluster of galaxies ($z=0.01724$) characterized by a
smooth distribution of ICM, and has a cD galaxy NGC~1129 at the
center.  In the central region within $2'$, Chandra observation showed
the temperature gradually dropped from 4 to 2 keV toward the cluster
center, and the metal abundance rose steeply to a peak of 1.5 solar
\citep{furusho03}.  \citet{ezawa97} also indicated a large-scale
abundance gradient of $\sim40\%$ from center to $r\sim 500$ kpc region
with ASCA\@.  \citet{hayakawa06} detected a temperature drop of
$\sim10\%$ from the central region to $r\sim13'$ with XMM-Newton,
while it had previously been considered to be flat based on ASCA
observation \citep{furusho01,ezawa97}.

This paper reports results from Suzaku observations of AWM~7
out to $27'\simeq 570\; h_{70}^{-1}$~kpc corresponding to
$\sim 0.35\; r_{180}$.  We use $H_0=70$
km~s$^{-1}$~Mpc$^{-1}$, $\Omega_{\Lambda} = 1-\Omega_M = 0.73$ in this
paper.  At the redshift of $z=0.01724$, $1'$ corresponds to 21~kpc,
and the virial radius, $r_{\rm 180} = 1.95\;
h_{100}^{-1}\sqrt{k\langle T\rangle/10~{\rm keV}}$~Mpc
\citep{markevitch98}, is 1.65~Mpc ($79'$) for the average temperature
of $k\langle T\rangle = 3.5$~keV\@.  Throughout this paper we adopt
the Galactic hydrogen column density of $N_{\rm H} = 9.83\times
10^{20}$ cm$^{-2}$ \citep{dickey90} in the direction of AWM~7\@.
Unless noted otherwise, the solar abundance table is given by
\citet{anders89}, and errors are 90\% confidence region for a single
interesting parameter.

\section{Suzaku Observation and Data Reduction}\label{sec:obs}

\begin{table*}
\caption{Suzaku observation of AWM~7}
\label{tab:1}
\begin{tabular}{lccccc} 
\hline \hline
Target name & Sequence number & Date & Exposure time & (RA, Dec) in J2000 $^\ast$ & After screening \\ 
\hline
AWM~7 center & 801035010 & 2006-Aug-7 & 19.0~ks & (\timeform{2h54m32.0s}, \timeform{+41D35'15''}) & 18.9~ks\\
AWM~7 east & 801036010 & 2006-Aug-5 & 38.5~ks & (\timeform{2h56m08.4s}, \timeform{+41D35'18''}) & 38.3~ks\\
AWM~7 west & 801037010 & 2006-Aug-6 & 39.8~ks & (\timeform{2h52m55.8s}, \timeform{+41D35'16''}) & 39.7~ks\\
\hline
\hline\\[-1ex]
\multicolumn{6}{l}{\parbox{0.95\textwidth}{\footnotesize 
\footnotemark[$\ast$]
Average pointing direction of the XIS, written in the 
RA\_NOM and DEC\_NOM keywords of the event FITS files.}}\\
\end{tabular}
\end{table*}

\subsection{Observation}
\label{subsec:obs}

Suzaku carried out three pointing observations for AWM~7 in August
2006 (PI:~T. Ohashi), for the central region and 20$'$-east and
$20'$-west offset regions with exposures of 19.0, 38.5 and 39.8~ks,
respectively.  The observation log is shown in table~\ref{tab:1}, and
the combined X-ray Imaging Spectrometer (XIS; \cite{koyama07}) image
in 0.5--7~keV range is shown in figure~\ref{fig:1}.  We analyze only
the XIS data in this paper.  The XIS instrument consists of four sets
of X-ray CCD (XIS~0, 1, 2, and 3). XIS~1 is a back-illuminated (BI)
sensor, while XIS~0, 2, and 3 are front-illuminated (FI) ones.  The
XIS was operated in the Normal clocking mode (8~s exposure per frame),
with the standard $5\times 5$ or $3\times 3$ editing mode.

It is known that the optical blocking filters (OBF) of the XIS have
been gradually contaminated by outgassing from the satellite.  The
thickness of the contaminant is different among the sensors, and is
also dependent on the location on the CCD chips.  The estimated column
density (C/O=6 in number ratio is assumed) during the observation at
the center of the CCD is listed in table~\ref{tab:2}.
\footnote{ Calibration database file of {\tt ae\_xi{\it
N}\_contami\_20061024.fits} was used for the estimation of the XIS
contamination ($N=0,1,2,3$ indicates the XIS sensor).}  We included
these effects in the calculation of the the Ancillary Response File
(ARF) by the ``xissimarfgen'' Ftools task of 2006-10-26 version
\citep{ishisaki07}.  Since the energy resolution also degraded slowly
after the launch due to the radiation damage, this effect was included
in the Redistribution Matrix File (RMF) by the ``xisrmfgen'' Ftools
task of 2006-10-26 version.

\begin{table}
\caption{Estimated column density of the contaminant for each sensor at 
the center of CCD in unit of 10$^{18}$~cm$^{-2}$.
}\label{tab:2}
\begin{center}
\begin{tabular}{cllll} 
\hline \hline
\hspace*{8em} &XIS0 & XIS1 & XIS2 &XIS3\\
\hline
Carbon $\dotfill$ & 2.53   & 3.97  & 3.83  & 5.79  \\
Oxygen $\dotfill$ & 0.422  & 0.662 & 0.639 & 0.964 \\
\hline 
\end{tabular}
\end{center}
\end{table}

\subsection{Data Reduction}

\begin{table}[t]
\caption{The best-fit parameters of figure~\ref{fig:3}.
}\label{tab:3}
\begin{center}
\begin{tabular}{cllc}
\hline\hline
\hspace*{6em} & \multicolumn{1}{c}{$\beta$} & \multicolumn{1}{c}{$r_{\rm c}$} &$S_{\makebox{\tiny 0.8--3~keV}}\,^*$ \\
\hline
narrower $\dotfill$ & $0.44\pm0.01$& $0.51'\pm 0.02'$ & $9.01\pm 0.28$ \\
wider $\dotfill$ & $0.60\pm0.01$&$5.06'\pm0.16'$ & $2.26\pm 0.08$\\
\hline\\[-1ex]
\multicolumn{4}{l}{\parbox{0.45\textwidth}{\footnotesize
\footnotemark[*]
$S_{\makebox{\tiny 0.8--3~keV}}$ at the center in unit of $10^{-7}$ counts~s$^{-1}$~arcsec$^{-2}$.}}\\
\end{tabular}
\end{center}
\end{table}

We used the version 1.2 processing data \citep{mitsuda07},
and the analysis was performed with HEAsoft version 6.1.1
and XSPEC 11.3.2t. 
The analysis method was almost same as \citet{sato07}. 
However, because this observations was not supported by the Good-Time 
Intervals (GTI) given for excluding the telemetry saturation 
by the XIS team, we could not execute the process.
The light curve of each sensor in the 0.3--10~keV range
with 16~s time bin was also examined to reject periods of
anomalous event rate greater or less than $\pm 3\sigma$ around the mean.
After the above screenings, remaining exposures of 
the observations were almost unchanged as shown in table~\ref{tab:1}.
The exposures after screening are not so different from 
those before screening in table~\ref{tab:1},
which represent that the non X-ray background (NXB) was 
almost stable during the both observations.
The event screening with the cut-off rigidity (COR)
was not performed in our data.

In order to subtract the NXB and the extra-galactic
cosmic X-ray background (CXB), we used dark earth
database of 770~ks exposure provided by the XIS team for the NXB, and
employed the CXB spectrum given by \citet{kushino02}.  These analysis
methods are also the same as in \citet{sato07}.

\begin{figure}[t]
\centerline{\FigureFile(0.45\textwidth,8cm){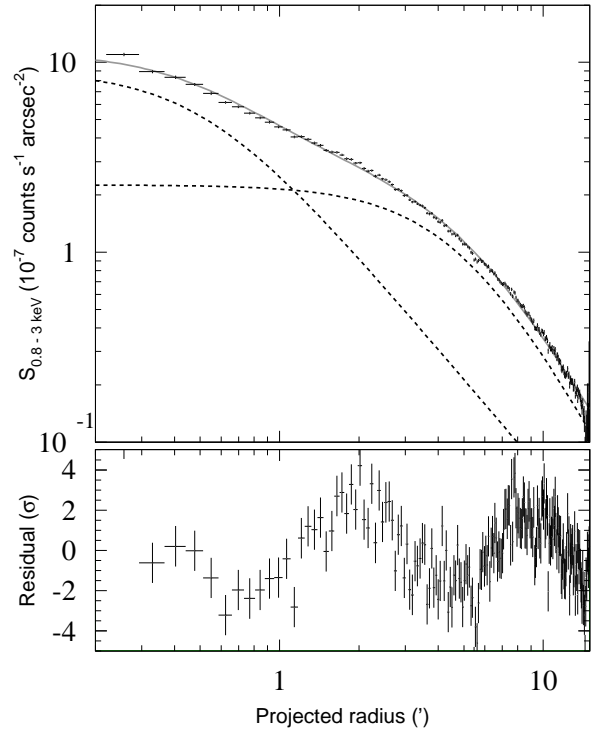}}
\vspace*{-1ex}
\caption{
In the upper panel, a radial profile of the surface brightness 
of AWM~7 in the 0.8--3~keV band are plotted for XMM-Newton MOS1+2 ($r<13'$). 
The best-fit double-$\beta$ model is shown by the solid gray line, and 
the two $\beta$-components are indicated by dashed lines.
In the bottom panel, fit residuals are shown in unit of $\sigma$, 
which correspond to the data minus the folded model divided
by the 1$\sigma$ error of each data point.
}\label{fig:2}
\end{figure}

\subsection{Generation of ARFs}
Precise surface brightness profile of AWM~7 was needed to generate the
Suzaku ARF, and we used the XMM-Newton image which had much better
spatial resolution than Suzaku.  For this XMM-Newton data, we used MOS
data (32 ks) and followed the data reduction by \citet{hayakawa06}.
We used the blank sky data as the background \citep{read03}, and
eliminated point sources in the ICM\@.  Figure~\ref{fig:2} shows a
radial profile of AWM~7 with XMM-Newton in the 0.8--3~keV range,
fitted with a double-$\beta$ model.  The origin of the profile is
placed at (RA, Dec) = (\timeform{2h54m27.5s}, \timeform{-41D34'46''})
in J2000.  The best-fit parameters are summarized in
table~\ref{tab:3}.  Though the fit is not acceptable, the temperature
and abundance in the spectral fits make little influence on the ARF\@.
Then, we generated two different ARFs for the spectrum of each region,
$A^{\makebox{\small\sc u}}$ and $A^{\makebox{\small\sc b}}$, which
respectively assume the uniform sky emission and $\sim 1^{\circ}
\times 1^{\circ}$ size of the double-$\beta$ surface brightness
profile obtained with the XMM-Newton data (table~\ref{tab:3}).  We
used not the raw XMM-Newton image but the smoothed surface brightness
profile to generate the ARFs, because the raw XMM image had the gap
between the CCD chips and AWM~7 was characterized by a smooth and
symmetric ICM distribution.

\section{Temperature \& Abundance profiles}\label{sec:spec}

\subsection{Spectral fit}
\label{sec:icm}
We extracted spectra from seven annular regions of 0--2$'$, 2--4$'$,
4--6$'$, 6--9$'$, 9--13$'$, 13--17$'$ and 17--27$'$, centered on (RA,
Dec) = (\timeform{2h54m32.0s}, \timeform{+41D35'15''}).  
The inner four annuli were
taken from the central observation, and the outer three annuli were
from the two offset observations.  Table~\ref{tab:4} lists areas of
the extraction regions (arcmin$^2$), coverage of the whole annulus
(\%), the {\sc source\_ratio\_reg} values
\footnote{{\sc source\_ ratio\_reg} represents flux ratio in the
assumed spatial distribution on the sky (double-$\beta$ model) inside
the accumulation region to the entire model, and written in the header
keyword of the calculated ARF by ``xissimarfgen''.}  and the observed
counts in the range 0.4--8.1~keV including NXB and CXB for the BI and
FI sensors.

\begin{table}[t]
\caption{
Area, coverage of whole annulus, {\sc source\_ratio\_ reg}
and observed counts for each annular region of AWM~7.
}\label{tab:4}
\begin{center}
\begin{tabular}{lrrrcc} 
\hline \hline
\makebox[3em][l]{Region\,$^\ast$} & \multicolumn{1}{c}{Area\makebox[0in][l]{\,$^\dagger$}} & Coverage\makebox[0in][l]{\,$^\dagger$}\hspace*{-0.5em} & \makebox[4.2em][r]{\sc source\_\makebox[0in][l]{\,$^\ast$}}\hspace*{-1em} & \multicolumn{2}{c}{Counts\makebox[0in][l]{\,$^{\ddagger}$}} \\
& \makebox[2em][c]{(arcmin$^2$)} &      & \makebox[4.2em][r]{\sc ratio\_reg}\hspace*{-1em} & BI     & FI \\
\hline
Center & & & & & \\
0--2$'$   & 12.6 &100.0\%       & 12.0\% & 21,800 & 52,737 \\
2--4$'$   & 37.7 &100.0\%       & 19.1\% & 38,654 & 89,901 \\
4--6$'$   & 62.8 &100.0\%       & 15.8\% & 33,146 & 75,797 \\
6--9$'$   &136.1 & 96.3\%       & 15.2\% & 34,767 & 74,968 \\
[-1.5ex]
\multicolumn{6}{l}{\hspace*{-0.6em}$\dotfill$\hspace*{-0.6em}} \\
East & & & & &\\
9--13$'$  & 48.5 & 17.6\%       &  2.1\% & 12,755 & 26,352 \\
13--17$'$ & 76.3 & 20.2\%       &  1.7\% & 14,577 & 31,750 \\
17--27$'$ &176.7 & 12.8\%       &  1.6\% & 19,242 & 39,475 \\
\multicolumn{6}{l}{\hspace*{-0.6em}$\dotfill$\hspace*{-0.6em}} \\
West & & & & &\\
9--13$'$  & 46.3 & 16.7\%       &  2.1\% & 10,695 & 23,108 \\
13--17$'$ & 72.7 & 19.3\%       &  1.6\% & 11,591 & 23,745 \\
17--27$'$ &182.9 & 13.2\%       &  1.7\% & 16,295 & 36,026 \\
\hline\\[-1ex]
\multicolumn{6}{l}{\parbox{0.45\textwidth}{\footnotesize
\footnotemark[$\ast$]
The inner four annuli are extracted from the central observation,
and the outer annuli are from the east and west observation, 
respectively.}} \\[1.2ex]
\multicolumn{6}{l}{\parbox{0.45\textwidth}{\footnotesize
\footnotemark[$\dagger$]
The largest values among four sensors are presented.}} \\
\multicolumn{6}{l}{\parbox{0.45\textwidth}{\footnotesize
\footnotemark[$\ddagger$]
Observed counts including NXB and CXB in 0.4--7.1 keV for BI and 
0.4--8.1 keV for FI.}}
\end{tabular}
\end{center}
\end{table}

\begin{figure*}
\begin{minipage}{0.33\textwidth}
\FigureFile(\textwidth,\textwidth){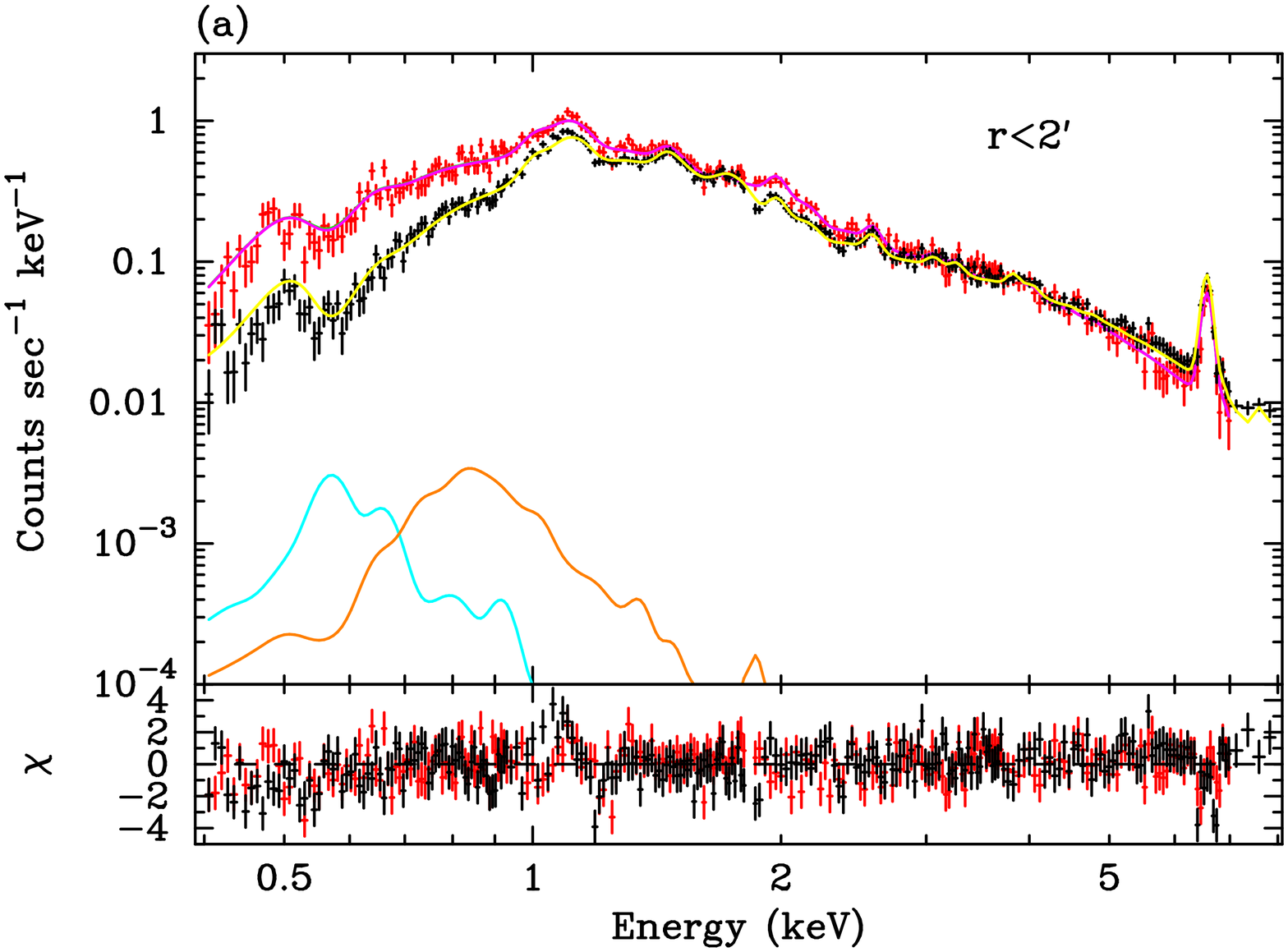}
\end{minipage}\hfill
\begin{minipage}{0.33\textwidth}
\FigureFile(\textwidth,\textwidth){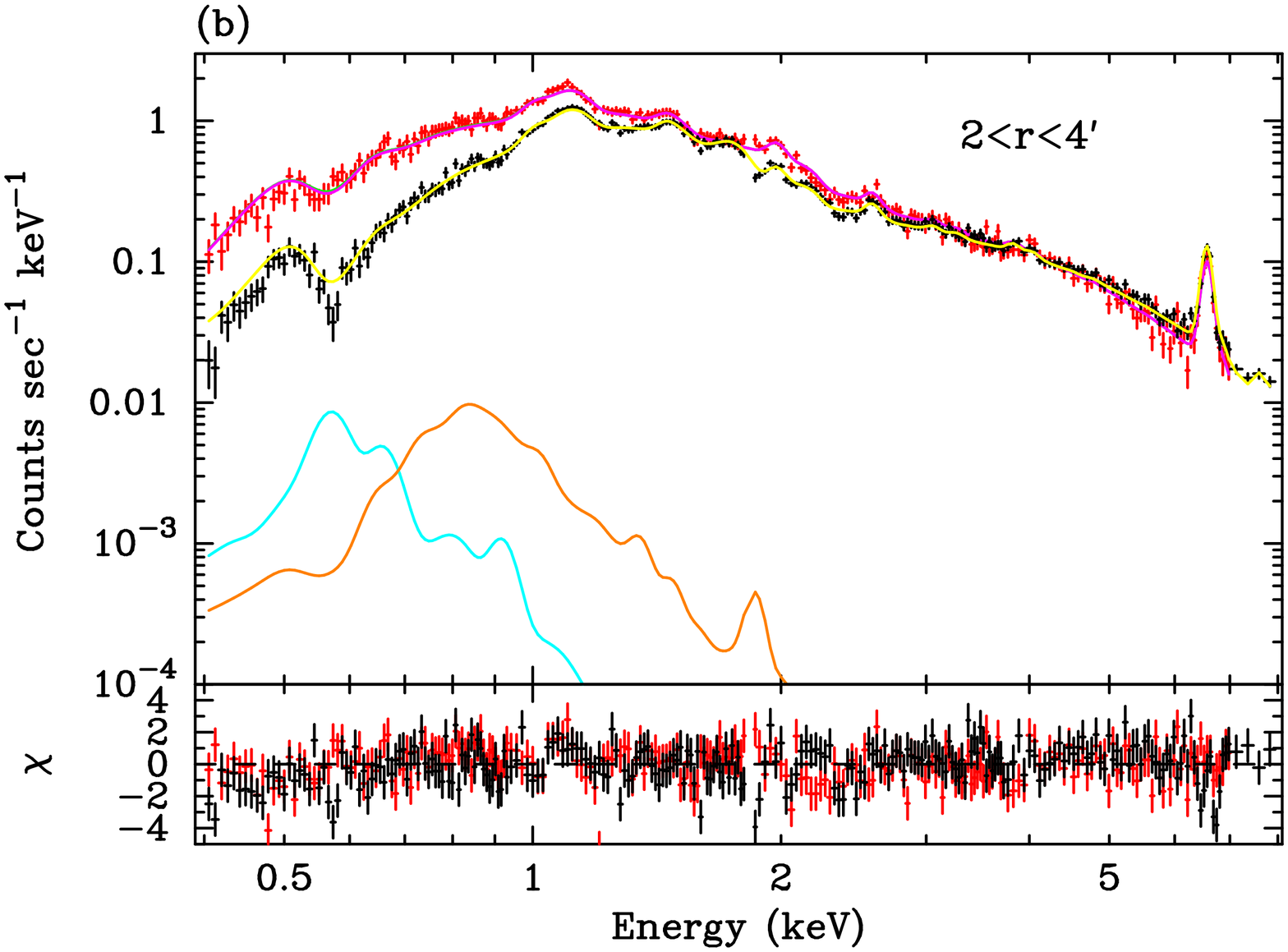}
\end{minipage}\hfill
\begin{minipage}{0.33\textwidth}
\FigureFile(\textwidth,\textwidth){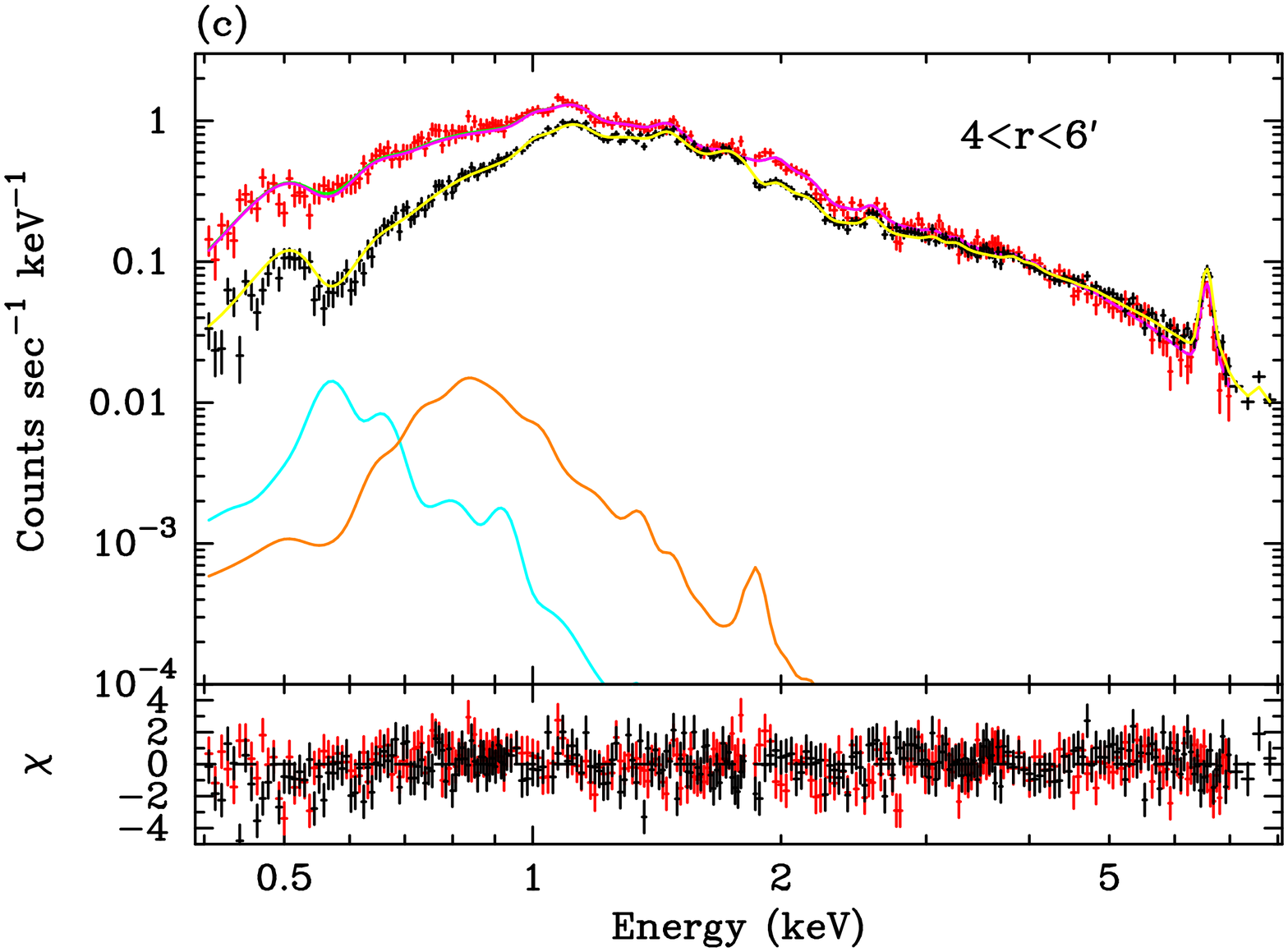}
\end{minipage}

\begin{minipage}{0.33\textwidth}
\FigureFile(\textwidth,\textwidth){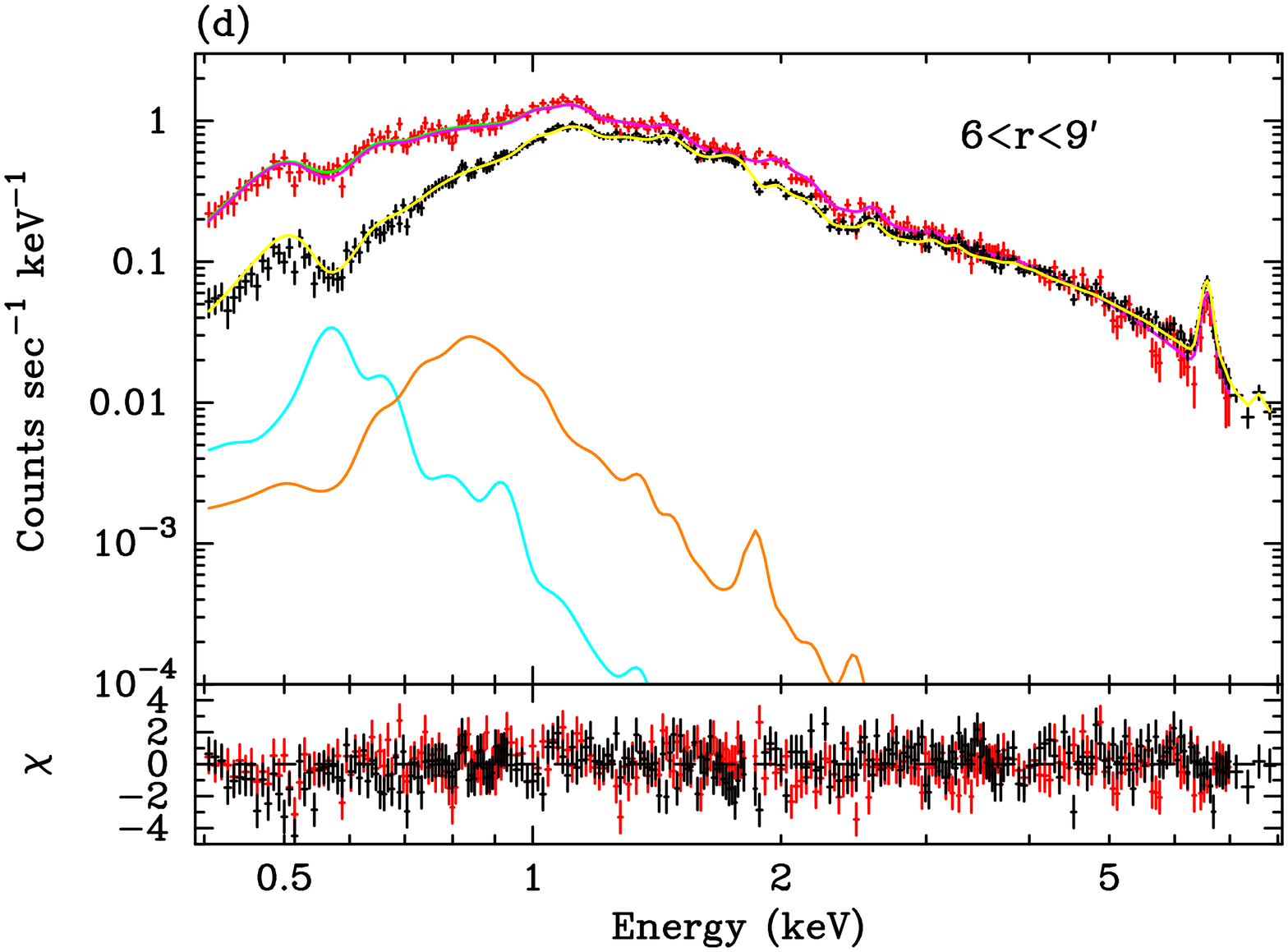}
\end{minipage}\hfill
\begin{minipage}{0.33\textwidth}
\FigureFile(\textwidth,\textwidth){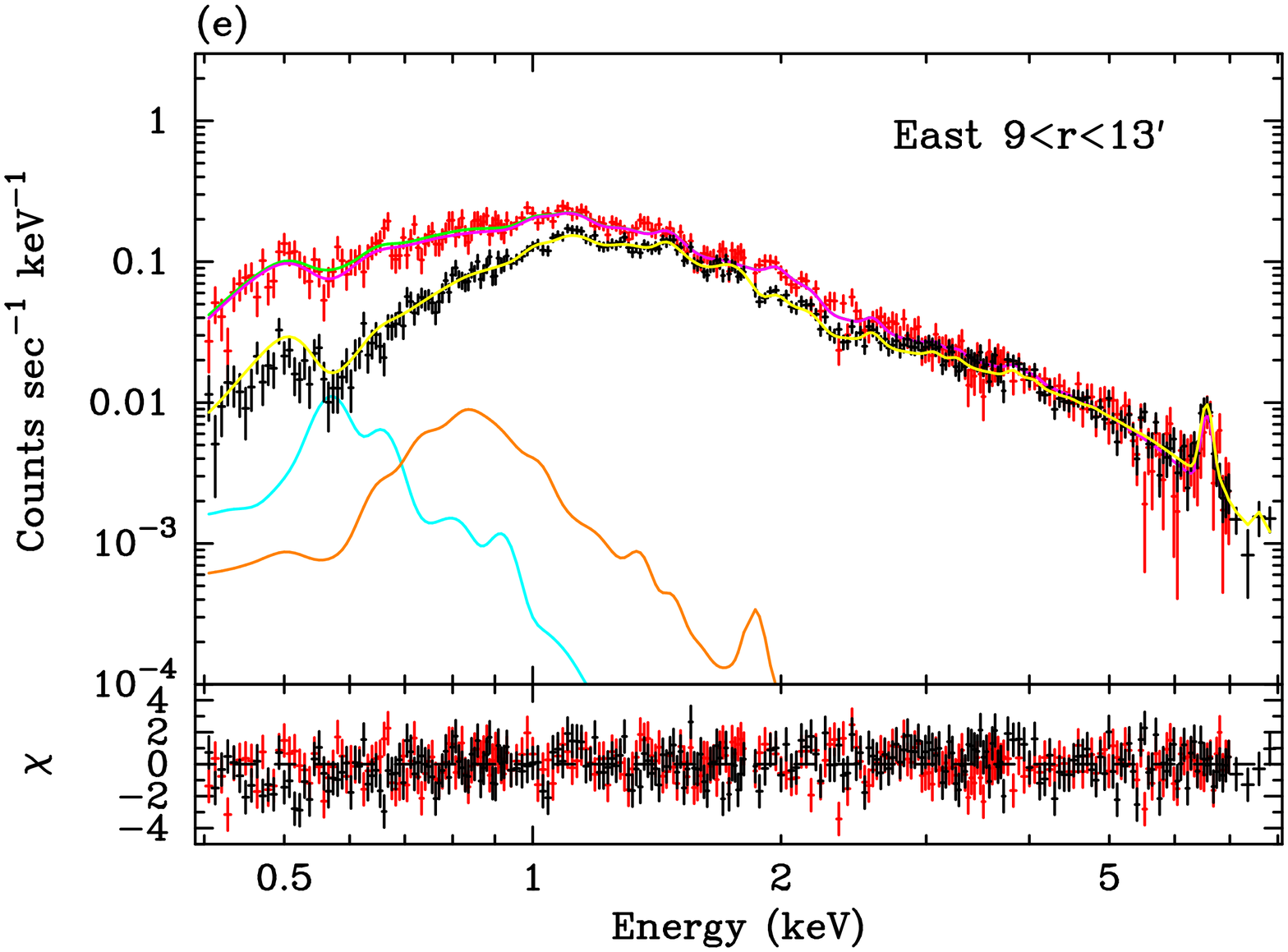}
\end{minipage}\hfill
\begin{minipage}{0.33\textwidth}
\FigureFile(\textwidth,\textwidth){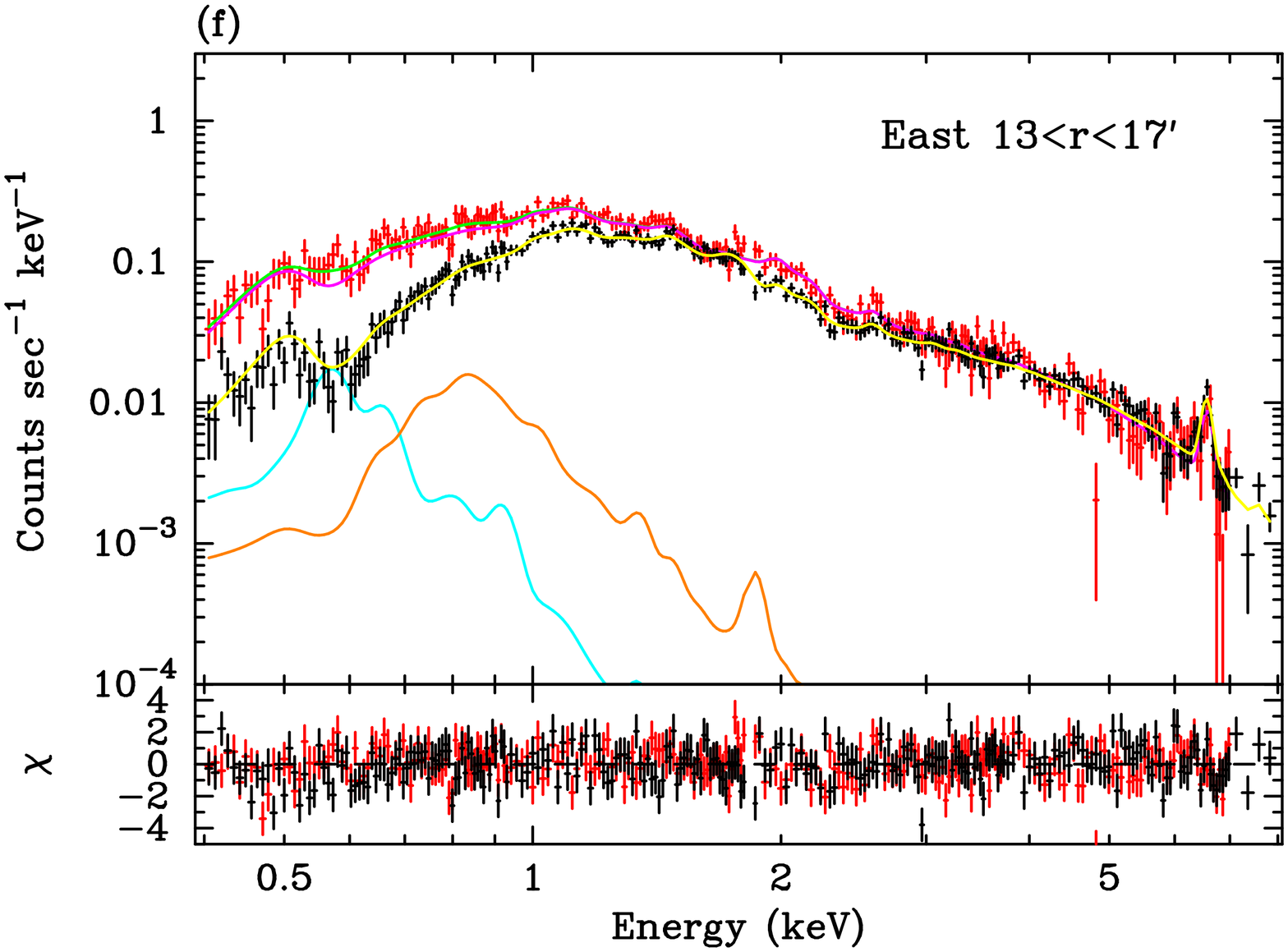}
\end{minipage}

\begin{minipage}{0.33\textwidth}
\FigureFile(\textwidth,\textwidth){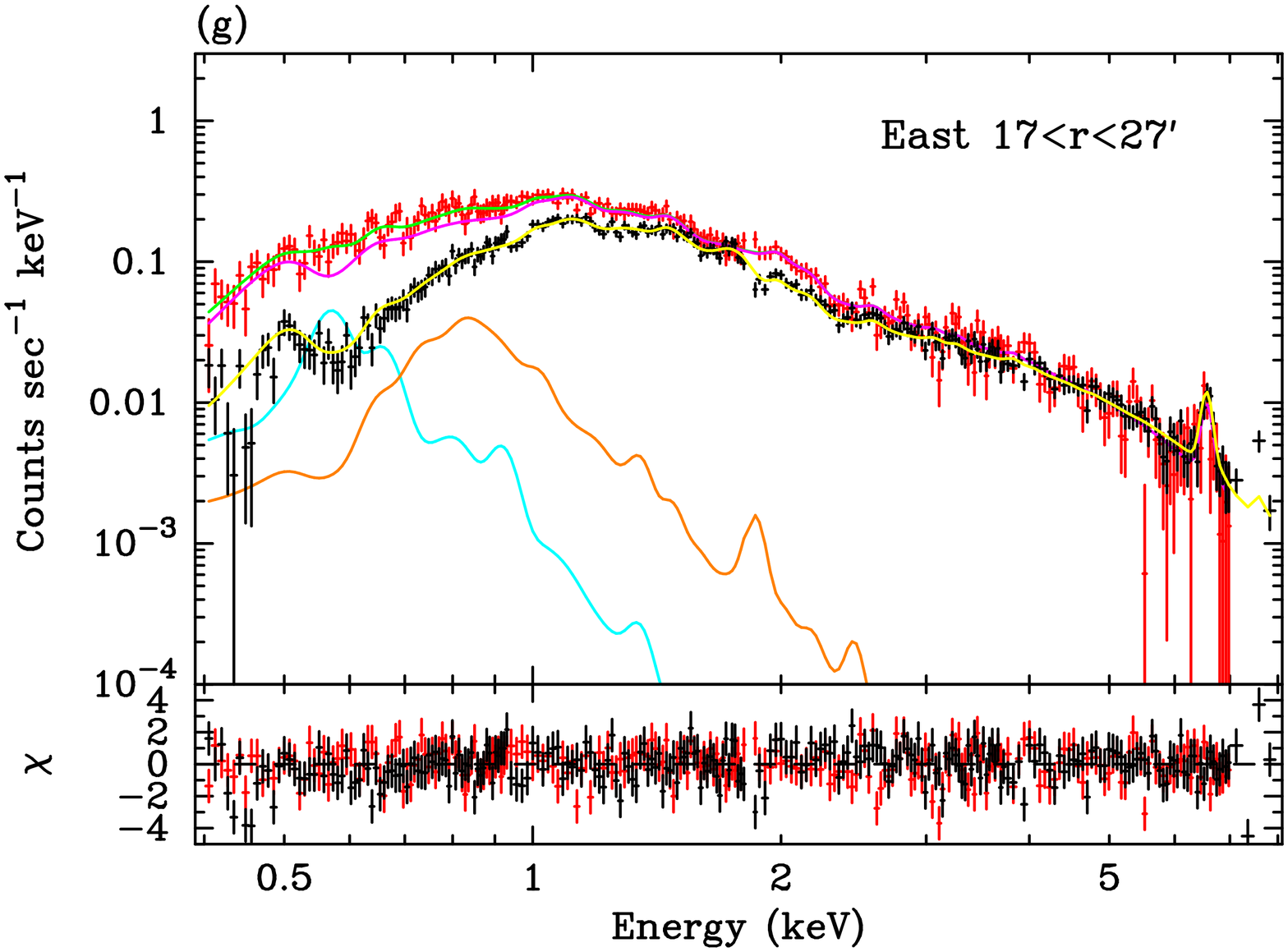}
\end{minipage}\hfill
\begin{minipage}{0.33\textwidth}
\FigureFile(\textwidth,\textwidth){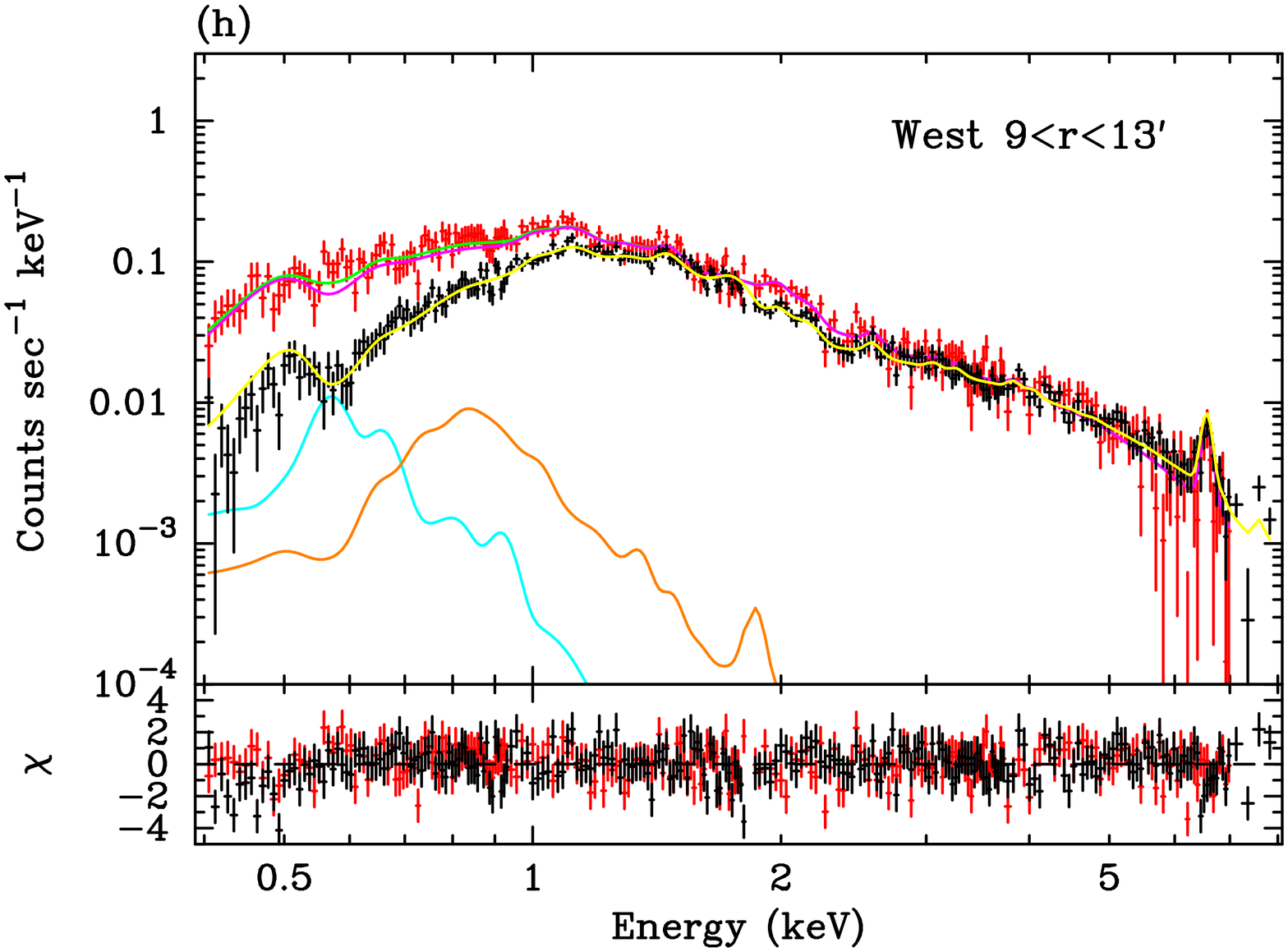}
\end{minipage}\hfill
\begin{minipage}{0.33\textwidth}
\FigureFile(\textwidth,\textwidth){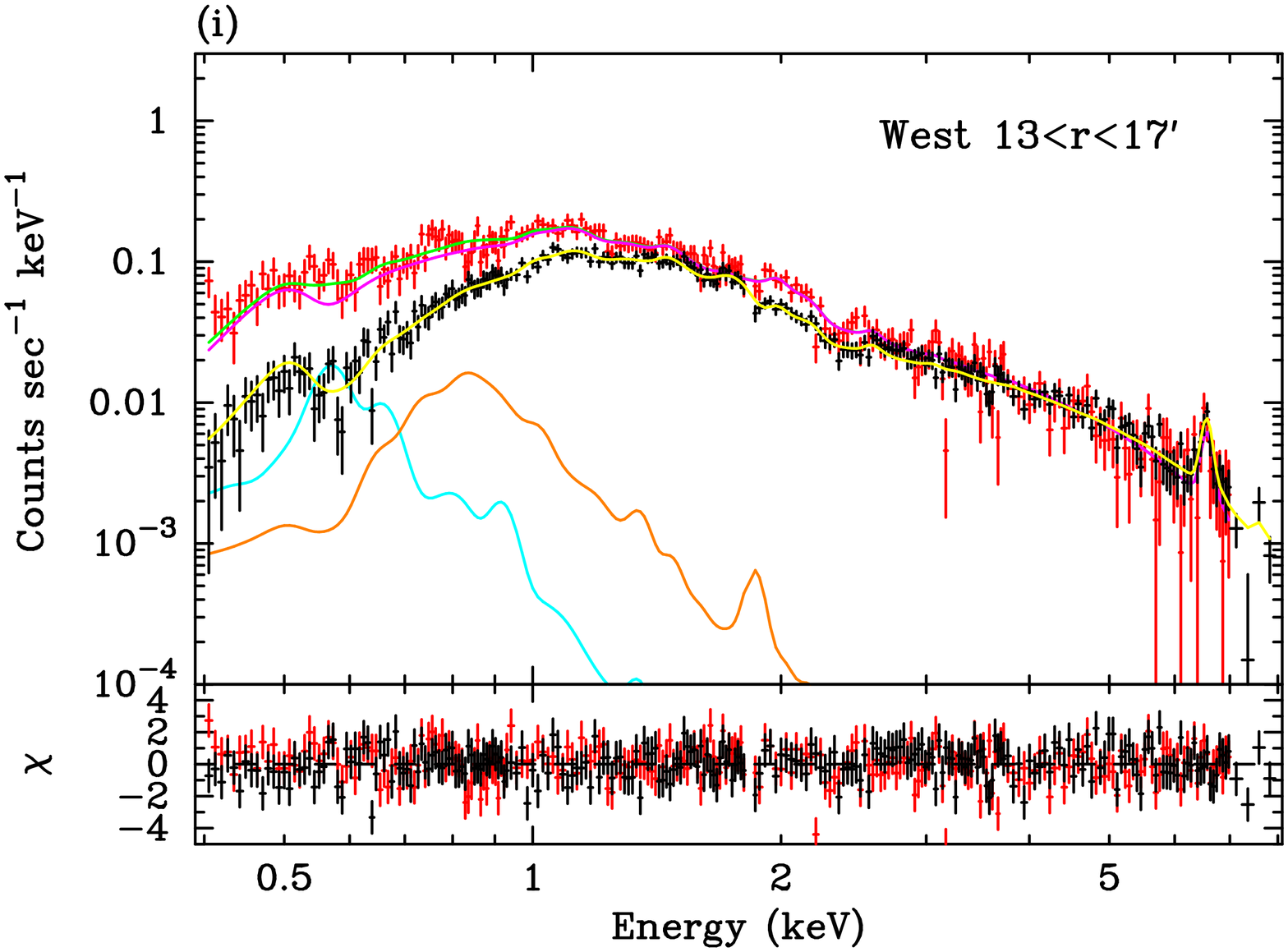}
\end{minipage}

\begin{minipage}{0.33\textwidth}
\FigureFile(\textwidth,\textwidth){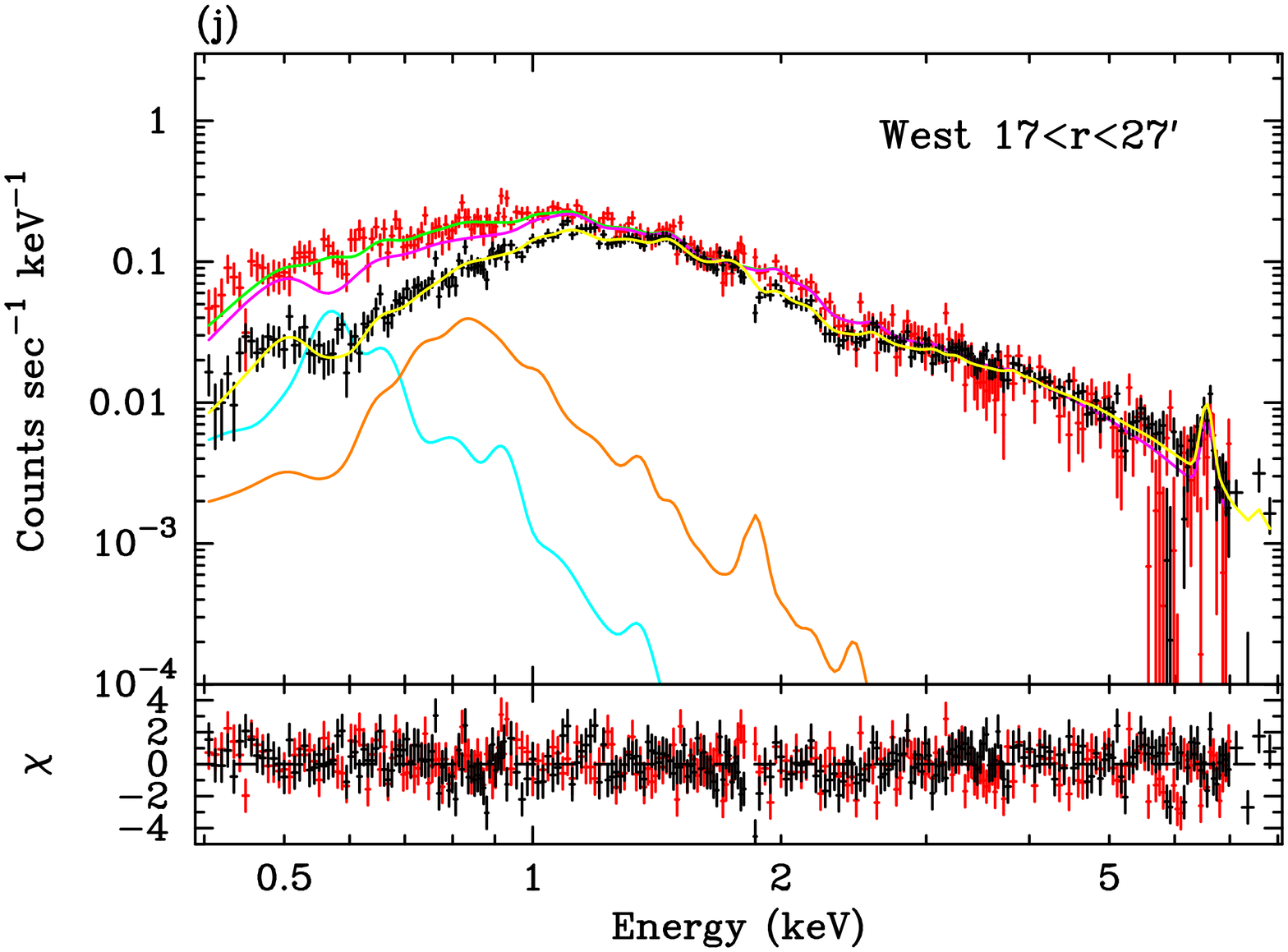}
\end{minipage}\hfill
\begin{minipage}{0.66\textwidth}
\caption{Upper panels show the observed spectra at the annular regions
as denoted in each panel, and red and black crosses are for BI and FI,
respectively.  The estimated CXB and NXB components are subtracted,
and the resultant spectra are fitted with ${\it apec}_1 + {\it
apec}_2 + {\it phabs} \times {\it vapec}$ model indicated by green and
yellow lines for the BI and FI spectra.  The ${\it apec}_1$ and ${\it
apec}_2$ components for the BI spectrum are indicated by cyan and
orange lines.  The energy range around the Si K-edge (1.825--1.840
keV) is ignored in the spectral fit.  The lower panels show the fit
residuals in unit of $\sigma$.  }
\label{fig:3}
\end{minipage}
\end{figure*}

The annular spectra are shown in figure~\ref{fig:3}.  Emission lines
from Mg, Si, S, Fe are clearly seen in the spectrum for each ring.
The O\emissiontype{VII} and O\emissiontype{VIII} lines are prominent
in the outer rings, however, most of the O\emissiontype{VII} emission
is considered to come from the local Galactic emission.

The basic strategy for the spectral fit is described in subsection 4.2
of \citet{sato07}.
The spectra for the east and west regions were fitted simultaneously
according to the distance from the cluster center.
The fit parameters were linked except for the normalization.
We will examine the difference between the east and west regions
in section 3.4.
The BI and FI spectra were also fitted simultaneously in the 0.4--8.1 keV
band excluding the response anomaly around Si K-edge
(1.825--1.840~keV)\@.  In the simultaneous fit, only the normalization
were allowed to take different values between BI and FI data,
though the derived normalizations were quite consistent between the two.

\subsection{Estimation of Galactic Component}
\label{subsec:galactic}

\begin{table*}
\caption{
The best-fit parameters of the {\it apec} component(s)
for the simultaneous fit of the spectra in 13--17$'$ and 17--27$'$ annuli 
of the west offset region and in 17--27$'$ annulus of the east offset region.
}\label{tab:5}
\begin{center}
\begin{tabular}{lccccc}
\hline\hline
\makebox[19em][l]{Fit model} & ${\it Norm}_1\,^\ast$ & $kT_1$ & ${\it Norm}_2\,^\ast$ & $kT_2$ & $\chi^2$/dof\\
 & & (keV) & & (keV) \\
\hline
(a) ${\it apec}_1 + {\it phabs}\times {\it vapec}$ $\dotfill$ &
        $7.52\pm 0.70$ & $0.262^{+0.035}_{-0.013}$ & --- & --- & 1781/1495\\
(b) ${\it apec}_1 + {\it apec}_2 + {\it phabs}\times {\it vapec}$ $\dotfill$ &
        $6.59\pm 0.46$ & $0.157^{+0.041}_{-0.035}$ & $0.32\pm 0.07$ & $0.612^{+0.052}_{-0.055}$ &  1743/1493\\
\hline\\[-1ex]

\multicolumn{6}{l}{\parbox{0.94\textwidth}{\footnotesize 
\footnotemark[$*$] 
Normalization of the {\it apec} component
divided by the solid angle, $\Omega^{\makebox{\tiny\sc u}}$,
assumed in the uniform-sky ARF calculation (20$'$ radius),
${\it Norm} = \int n_{\rm e} n_{\rm H} dV \,/\,
(4\pi\, (1+z)^2 D_{\rm A}^{\,2}) \,/\, \Omega^{\makebox{\tiny\sc u}}$
$\times 10^{-20}$ cm$^{-5}$~arcmin$^{-2}$,
where $D_{\rm A}$ is the angular distance to the source.}}
\end{tabular}
\end{center}
\end{table*}

It is important to estimate the foreground Galactic emission
precisely, for which the offset observations of AWM~7 give useful data.  The
Galactic component is prominent in the 17--27$'$ annulus as shown in
figure~\ref{fig:3}, however the ICM component is still dominant in
almost all the energy range except for the O\emissiontype{VII} line.
We also dealt with these data in the same way as in \citet{sato07}.  We
performed simultaneous fit in the whole 0.4--8.1~keV range (except
for 1.825--1.840~keV) for the spectra in 13--17$'$ annulus of the west
offset region and the one in 17--27$'$ annuli of the east and west offset
regions. We assumed either one or two temperature {\it apec} model for
the Galactic component, and the fit results are presented in
table~\ref{tab:5}.  The resultant normalizations of the {\it apec}
models in table~\ref{tab:5} are scaled so that they give the surface
brightness per arcmin$^2$.

We examined the improvement of $\chi^2$ ($\Delta\chi^2 = 38$) values
with the $F$-test, and addition of the ${\it apec}_2$ component was
necessary with high significance (error probability $\sim 10^{-7}$).
We further allowed the normalization of the ${\it apec}_2$ to be free
for those three annuli.  We concluded that the two {\it apec} models
are required to account for the Galactic component. Hereafter, the
model ${\it apec}_1 + {\it apec}_2 + {\it phabs}\times {\it vapec}$ is
used for the spectral fits, unless otherwise stated.

In order to take into account both the existence of the Galactic
component itself and the propagation of its statistical error, we
fitted the spectrum in each individual annulus simultaneously with the
one in the outermost annulus, i.e.\ at 17--27$'$ in the east and west
offset regions.  The normalization for the sum of the ${\it apec}_1$
and ${\it apec}_2$ components was constrained to give the same surface
brightness in all annuli, and the temperatures of these two {\it apec}
models were also common for all the regions including the 17--27$'$
annuli.  These common normalizations, ${\it Norm}_1$ and ${\it
Norm}_2$, and the common temperatures, $kT_1$ and $kT_2$, were the
free parameters.  The derived normalizations and temperatures for the
Galactic components are consistent with the values already shown in
table~\ref{tab:5}(b) within their errors.  Influence on the ICM
temperature and abundance due to the modeling of the Galactic
component will be examined in subsection \ref{subsec:radial}.

\subsection{Radial Temperature \& Abundance Profiles}
\label{subsec:radial}

\begin{table*}
\caption{
Summary of the best-fit parameters of the {\it vapec} component
for each annular region of AWM~7.
with the
${\it apec}_1 + {\it apec}_2 + {\it phabs} \times {\it vapec}$ model$^\ast$.
}\label{tab:6}
\centerline{
\begin{tabular}{lrccccccccc}
\hline\hline\\[-2ex]
\makebox[2em][l]{Region} & {\it Norm}\makebox[0in][l]{\,$^\dagger$} & $kT$&O&Ne&Mg&Si&\makebox[0in][c]{S}&Fe&Ni & $\chi^2$/dof \\
& &(keV)&(solar)&(solar)&(solar)&(solar)&(solar)&(solar)&(solar) & \\
\hline\\[-2ex]
0--2$'$    &  $1007\pm36$  & 3.41$^{+0.04}_{-0.04}$ & 0.62$^{+0.34}_{-0.31}$ & 2.11$^{+0.40}_{-0.37}$ & 1.50$^{+0.32}_{-0.31}$ & 1.13$^{+0.18}_{-0.17}$ & 1.08$^{+0.20}_{-0.20}$ & 0.87$^{+0.05}_{-0.04}$ & 1.71$^{+0.61}_{-0.58}$  &  1872/1493\\\\[-2ex]
2--4$'$    &  $595\pm16$  & 3.66$^{+0.04}_{-0.04}$ & 0.57$^{+0.26}_{-0.25}$ & 1.65$^{+0.28}_{-0.27}$ & 1.18$^{+0.24}_{-0.24}$ & 0.94$^{+0.13}_{-0.13}$ & 0.85$^{+0.16}_{-0.15}$ & 0.73$^{+0.03}_{-0.03}$ & 1.47$^{+0.47}_{-0.46}$  & 1899/1493\\\\[-2ex]
4--6$'$    &  $342\pm10$  & 3.78$^{+0.05}_{-0.05}$ & 0.49$^{+0.28}_{-0.26}$ & 1.38$^{+0.29}_{-0.28}$ & 1.20$^{+0.27}_{-0.27}$ & 0.54$^{+0.14}_{-0.14}$ & 0.67$^{+0.17}_{-0.17}$ & 0.56$^{+0.03}_{-0.03}$ & 0.56$^{+0.53}_{-0.52}$  & 1805/1493\\\\[-2ex]
6--9$'$    &  $182\pm5$  & 3.82$^{+0.06}_{-0.06}$ & 0.63$^{+0.25}_{-0.27}$ & 1.33$^{+0.28}_{-0.29}$ & 1.05$^{+0.26}_{-0.27}$ & 0.57$^{+0.14}_{-0.15}$  & 0.74$^{+0.18}_{-0.18}$ & 0.50$^{+0.03}_{-0.03}$ & 2.11$^{+0.57}_{-0.56}$  & 1816/1493\\\\[-2ex]
\hline
\makebox[0in][l]{9--13$'$}   &  $ 113\pm4$  & 3.62$^{+0.08}_{-0.08}$ & 0.39$^{+0.33}_{-0.34}$ & 1.08$^{+0.33}_{-0.32}$ & 1.32$^{+0.32}_{-0.33}$ & 0.53$^{+0.17}_{-0.17}$ & 0.62$^{+0.21}_{-0.21}$ & 0.41$^{+0.04}_{-0.04}$ & 2.33$^{+0.70}_{-0.68}$  & 2361/2014\\\\[-2ex]
\makebox[0in][l]{13--17$'$}  &  $ 72\pm2$  & 3.58$^{+0.08}_{-0.08}$ & 0.10$^{+0.37}_{-0.10}$ & 0.86$^{+0.30}_{-0.31}$ & 0.75$^{+0.29}_{-0.28}$ & 0.50$^{+0.16}_{-0.15}$ & 0.38$^{+0.19}_{-0.18}$ & 0.32$^{+0.03}_{-0.03}$ & 1.22$^{+0.64}_{-0.61}$  & 2370/2014\\\\[-2ex]
\makebox[0in][l]{17--27$'$}  &  $ 29\pm1$  & 3.34$^{+0.07}_{-0.07}$ & 0.46$^{+0.57}_{-0.46}$ & 0.91$^{+0.32}_{-0.31}$ & 0.88$^{+0.29}_{-0.27}$ & 0.42$^{+0.15}_{-0.14}$ & 0.29$^{+0.17}_{-0.17}$ & 0.36$^{+0.04}_{-0.04}$ & 1.88$^{+0.64}_{-0.60}$  &  ---\makebox[0in][l]{\,$^\ddagger$} \\\\[-2ex]
\hline
\\[-1ex]
\multicolumn{11}{l}{\parbox{0.96\textwidth}{\footnotesize
\footnotemark[$\ast$]
Spectrum in each annulus was simultaneously fitted with the outermost
17--27$'$ spectrum obtained by the two offset observations. Errors
show 90\% confidence statistical error range, without including
systematic errors.  These results are shown in figure~\ref{fig:4},
where Ne and Ni abundances may have large systematic uncertainties
because XIS cannot resolve the ionized Ne lines from the
Fe-L line complex.  }}\\
\multicolumn{11}{l}{\parbox{0.96\textwidth}{\footnotesize
\footnotemark[$\dagger$] 
Normalization of the {\it vapec} component scaled with a factor of
{\sc source\_ratio\_reg} / {\sc area} in table~\ref{tab:4},\\ ${\it
Norm}=\frac{\makebox{\sc source\_ratio\_reg}}{\makebox{\sc area}} \int
n_{\rm e} n_{\rm H} dV \,/\, (4\pi\, (1+z)^2 D_{\rm A}^{\,2})$ $\times
10^{-20}$~cm$^{-5}$~arcmin$^{-2}$, where $D_{\rm A}$ is the angular
distance to the source.  For the offset regions, these are the values
of normalization of the east region.}}\\
\multicolumn{11}{l}{\parbox{0.96\textwidth}{\footnotesize
\footnotemark[$\ddagger$] The 17--27$'$ annulus of the east and west
offset regions were fitted simultaneously with other annuli, and the
best-fit values with the 13--17$'$ annulus of the east and west offset
regions, respectively, are presented here.  }}
\end{tabular}}
\end{table*}

\begin{figure*}
\begin{minipage}{0.33\textwidth}
\FigureFile(\textwidth,\textwidth){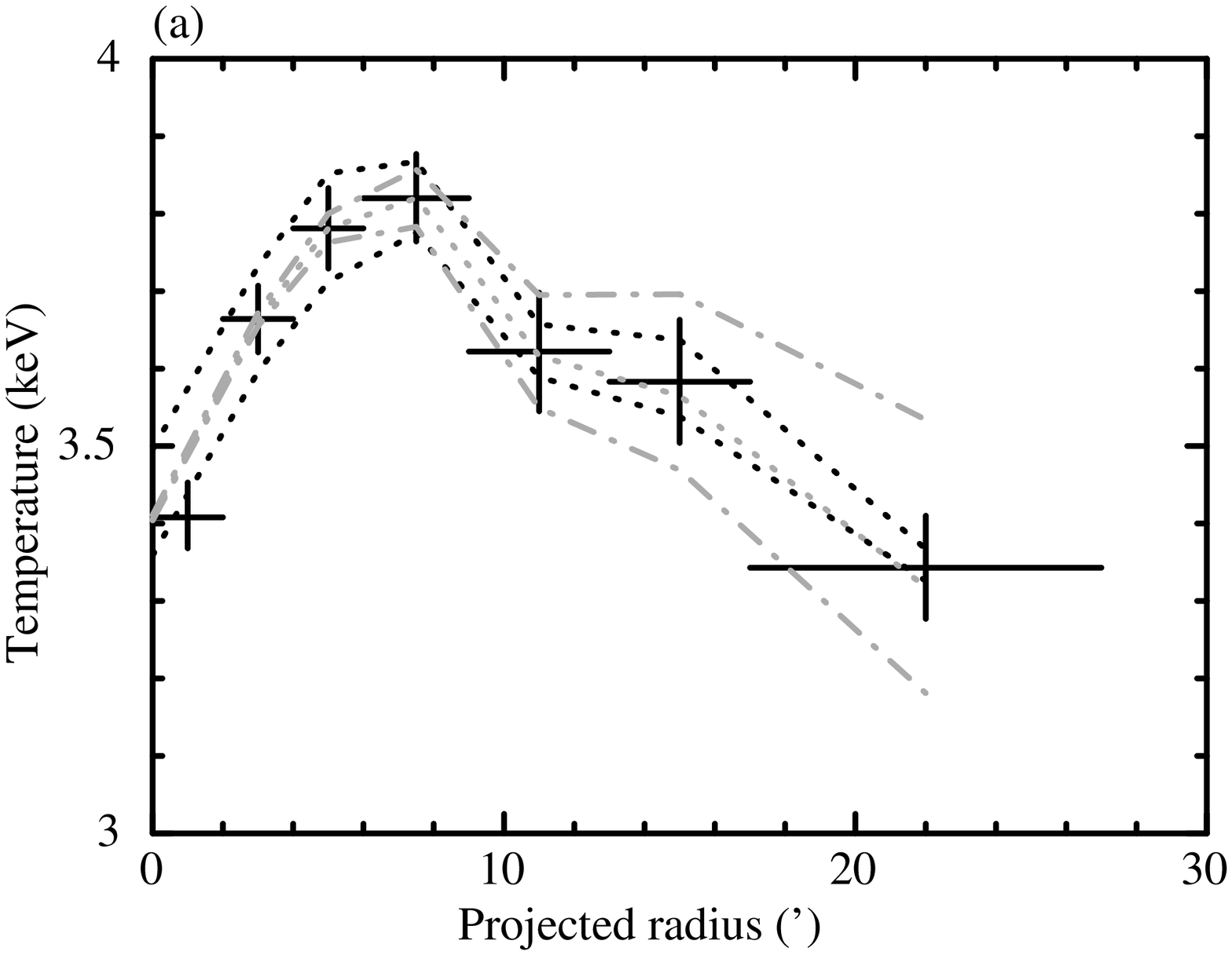}
\end{minipage}\hfill
\begin{minipage}{0.33\textwidth}
\FigureFile(\textwidth,\textwidth){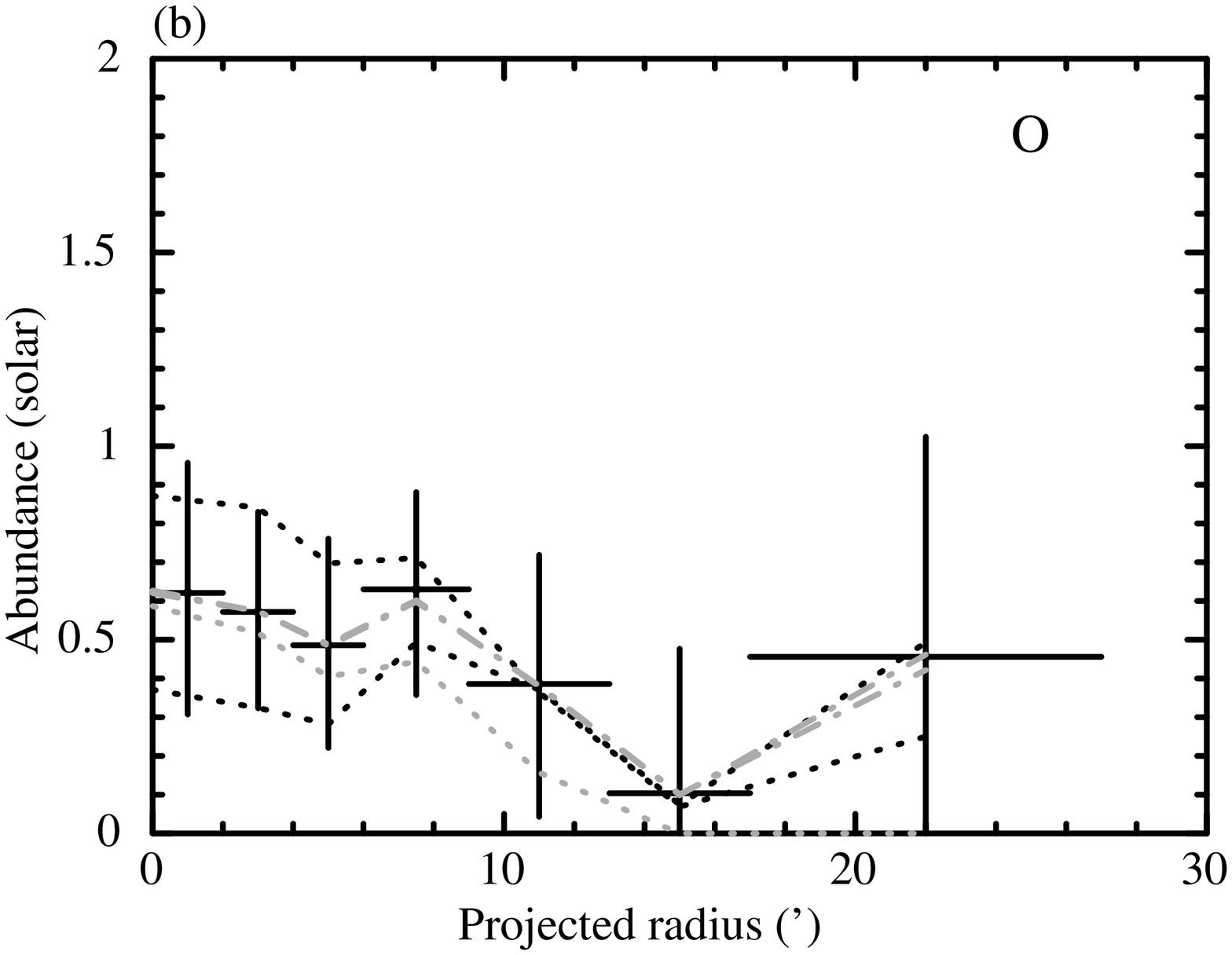}
\end{minipage}\hfill
\begin{minipage}{0.33\textwidth}
\FigureFile(\textwidth,\textwidth){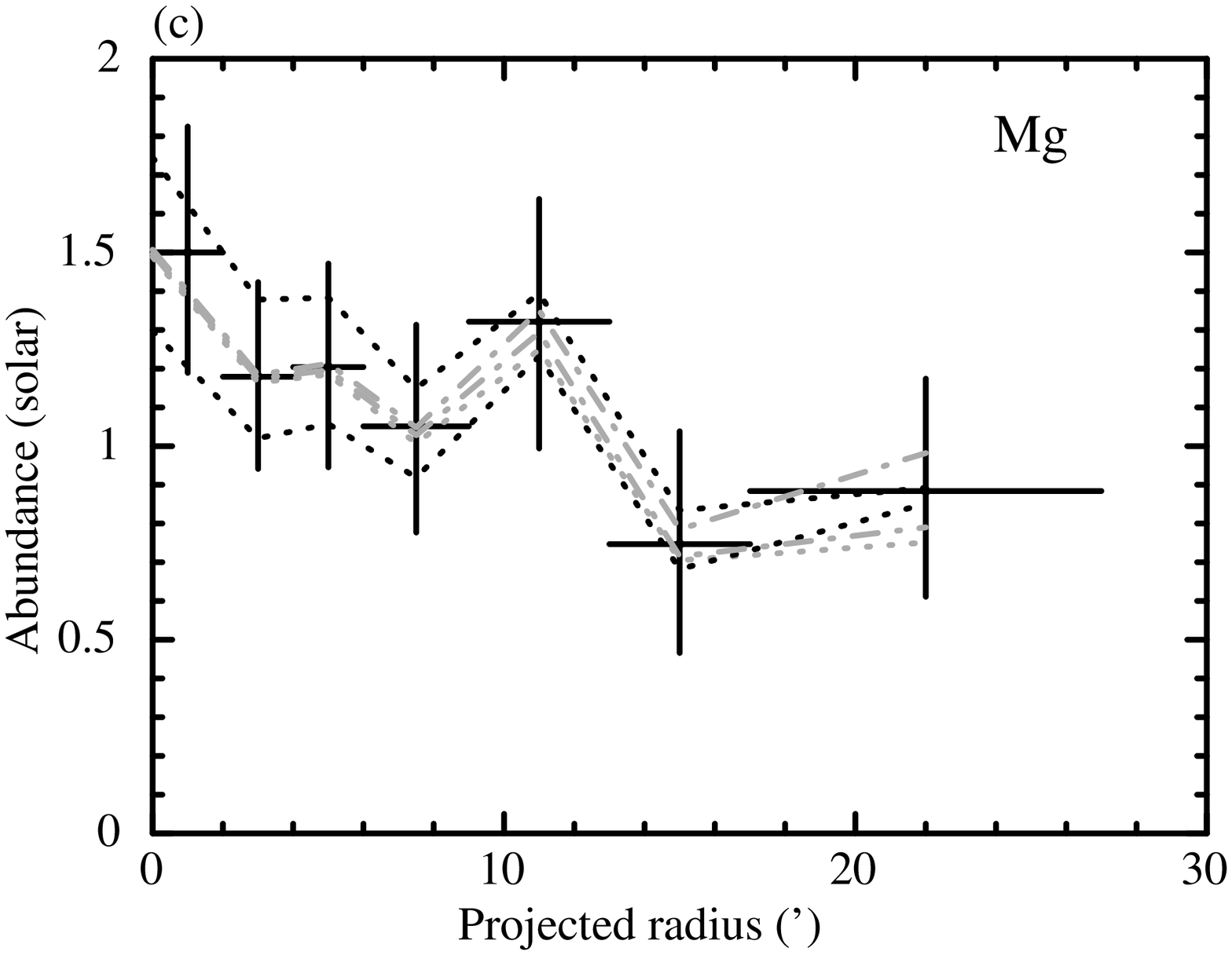}
\end{minipage}

\begin{minipage}{0.33\textwidth}
\FigureFile(\textwidth,\textwidth){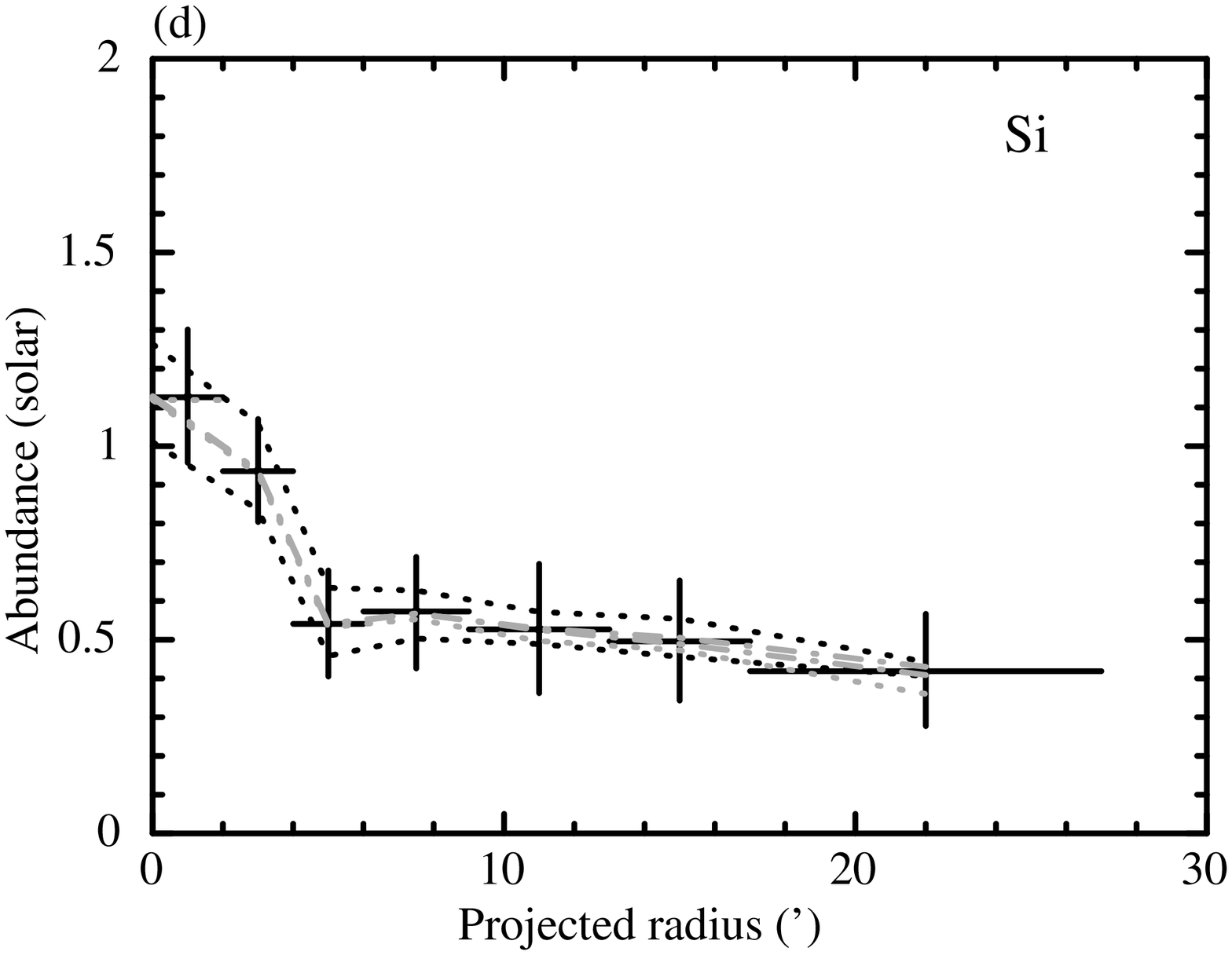}
\end{minipage}\hfill
\begin{minipage}{0.33\textwidth}
\FigureFile(\textwidth,\textwidth){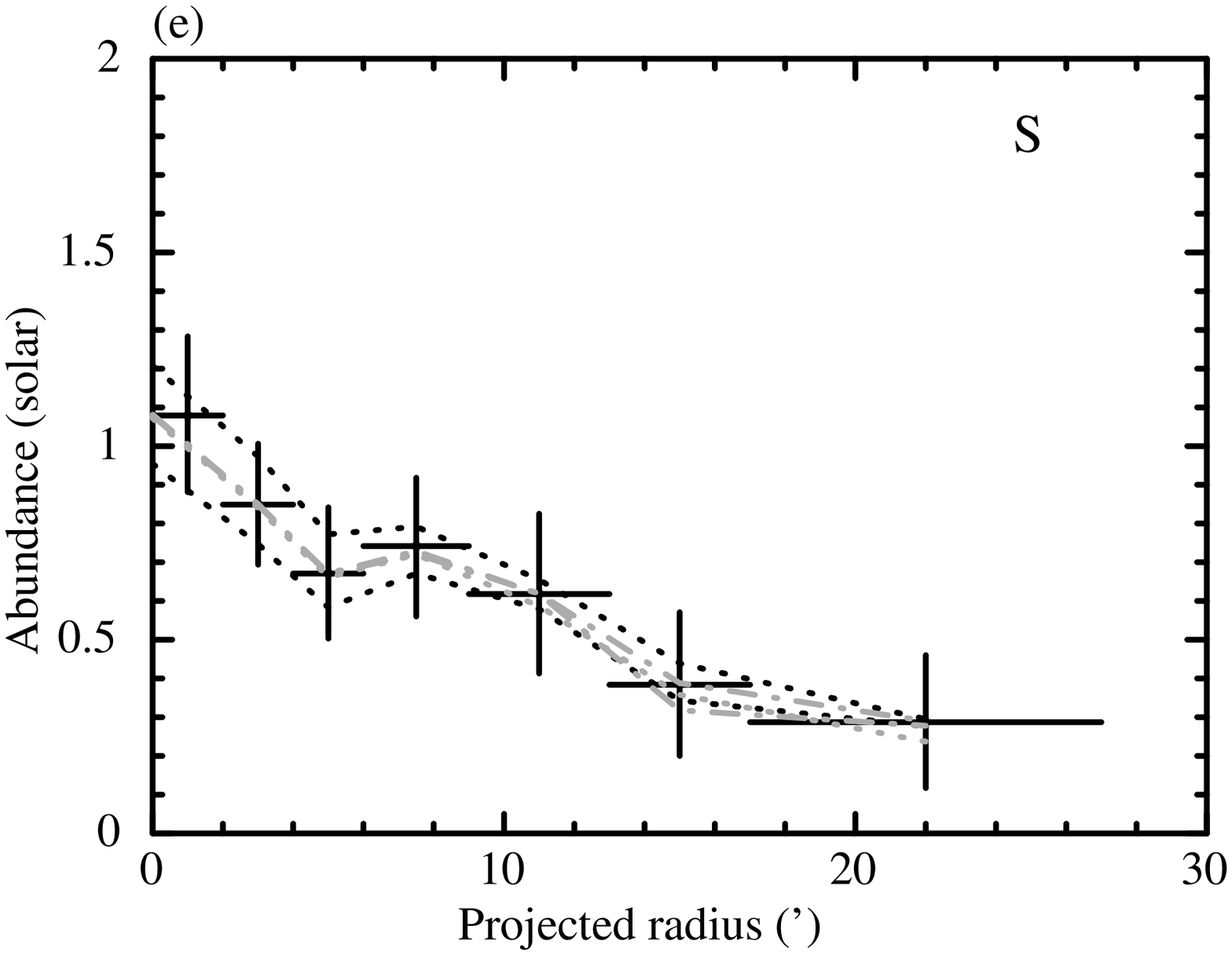}
\end{minipage}\hfill
\begin{minipage}{0.33\textwidth}
\FigureFile(\textwidth,\textwidth){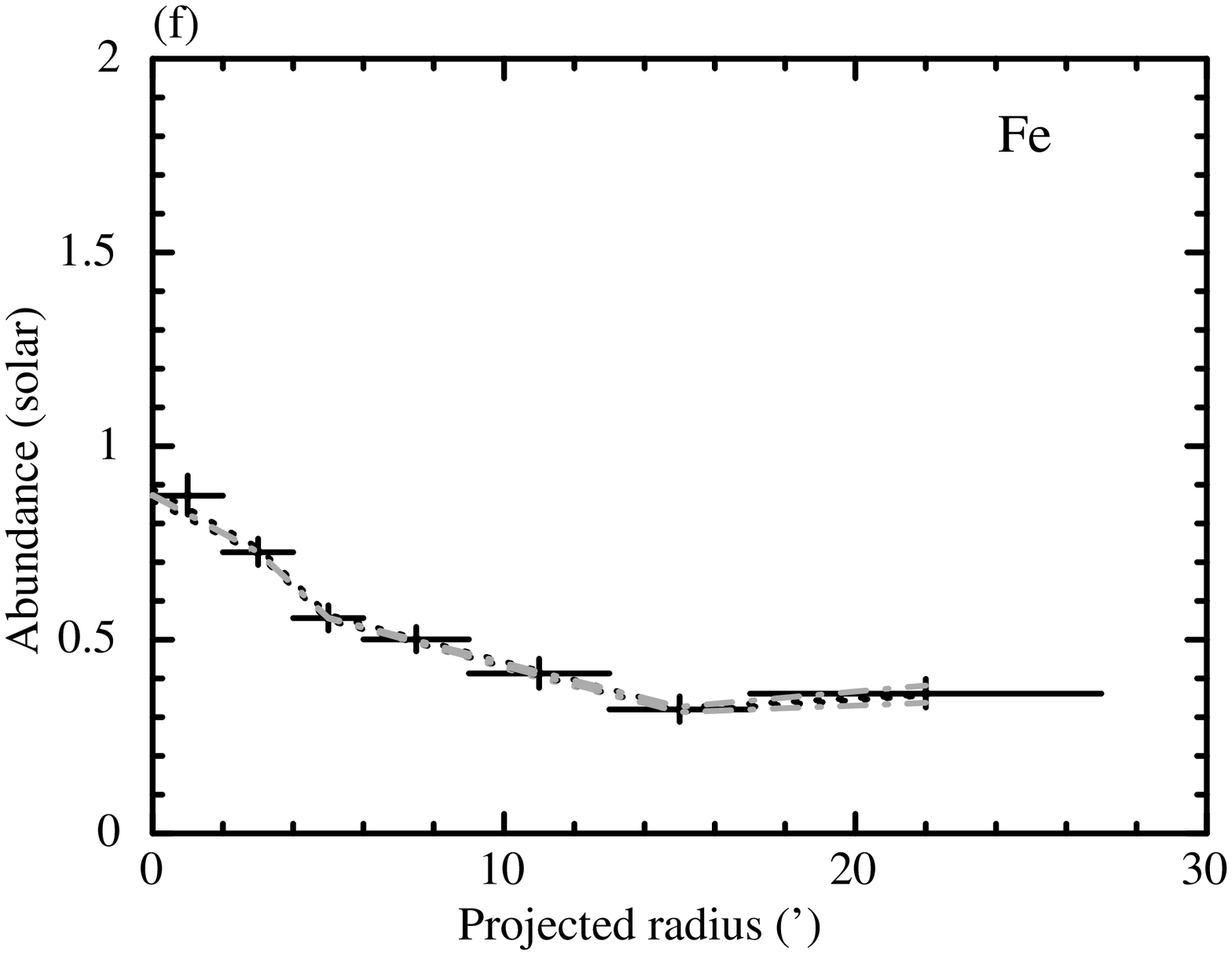}
\end{minipage}

\vspace*{-0.5ex}
\caption{(a) Radial temperature profiles derived from the spectral fit
for each annulus.  The horizontal axis denotes the projected radius.
Black dotted lines indicate shifts of the best-fit values by changing
thickness of the OBF contaminant by $\pm 10$\%.  Light-gray dotted lines
denote those when the estimated CXB and NXB levels are changed by $\pm
10$\%.  Light-gray  dash-dotted line shows the best-fit value when the Galactic
component is modeled by a single temperature {\it apec} model.
(b)--(f) Radial abundance profiles derived and plotted in the same way
as in (a).  }
\label{fig:4}
\end{figure*}

\begin{table*}
\caption{List of $\chi^2$/dof for each fit of AWM~7.}
\label{tab:7}
\begin{center}
\begin{tabular}{lccccc}
\hline\hline
\makebox[6em][l]{Region} & nominal &\multicolumn{2}{c}{contaminant} & \multicolumn{2}{c}{background}\\
\hline
 & & \makebox[0in][c]{+10\%} & \makebox[0in][c]{-10\%} & \makebox[0in][c]{+10\%} & \makebox[0in][c]{-10\%}\\
\hline
center $\dotfill$  &  &  &  &  & \\
0--2$'$ $\dotfill$   & 1872/1493 & 1829/1493 & 1939/1493 & 1909/1493 & 1871/1493 \\
2--4$'$ $\dotfill$   & 1899/1493 & 1876/1493 & 1965/1493 & 1937/1493 & 1897/1493 \\
4--6$'$ $\dotfill$   & 1805/1493 & 1789/1493 & 1849/1493 & 1844/1493 & 1803/1493 \\
6--9$'$ $\dotfill$   & 1816/1493 & 1786/1493 & 1852/1493 & 1858/1493 & 1814/1493 \\
\hline
East \& West $\dotfill$  &  &  &  &  & \\
9--13$'$ $\dotfill$  & 2361/2014 & 2337/2014 & 2395/2014 & 2408/2014 & 2355/2014 \\
13--17$'$ $\dotfill$ & 2370/2014 & 2365/2014 & 2390/2014 & 2424/2014 & 2362/2014 \\
\hline
\end{tabular}
\end{center}
\end{table*}

\begin{figure*}
\begin{center}
\begin{minipage}{0.45\textwidth}
\centerline{\FigureFile(\textwidth,\textwidth){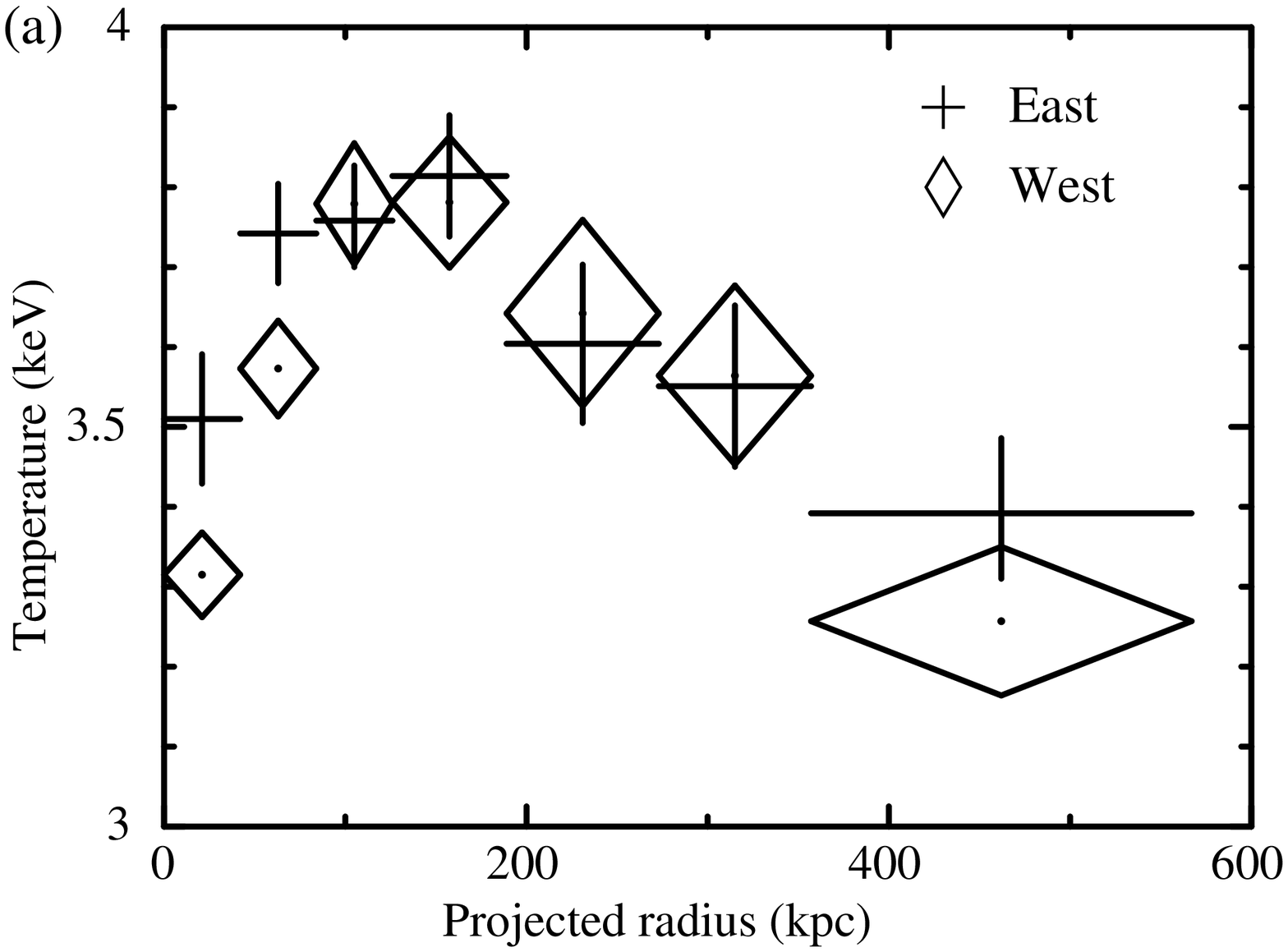}}
\end{minipage}
\hfill
\begin{minipage}{0.45\textwidth}
\centerline{\FigureFile(\textwidth,\textwidth){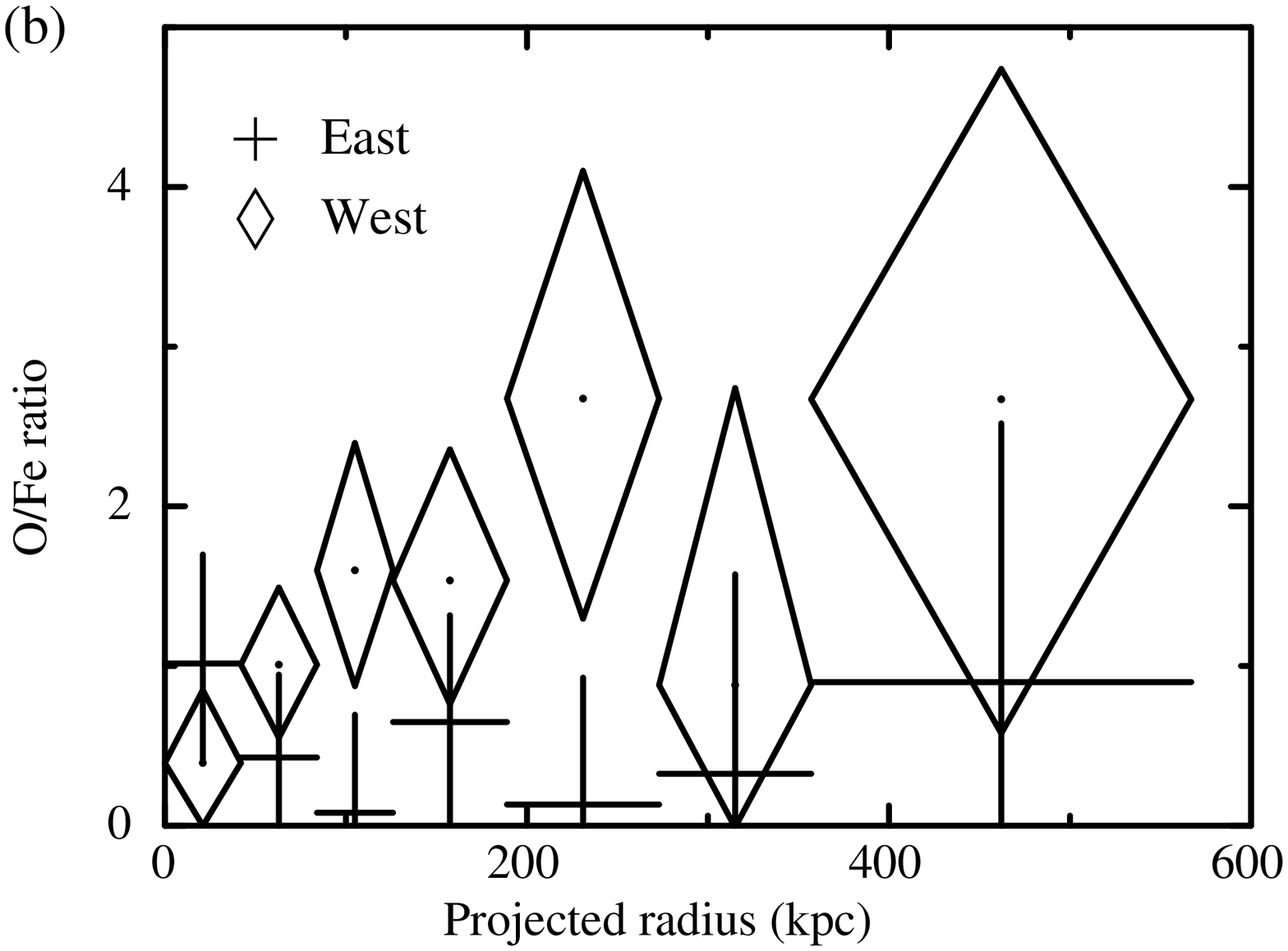}}
\end{minipage}
\begin{minipage}{0.45\textwidth}
\centerline{\FigureFile(\textwidth,\textwidth){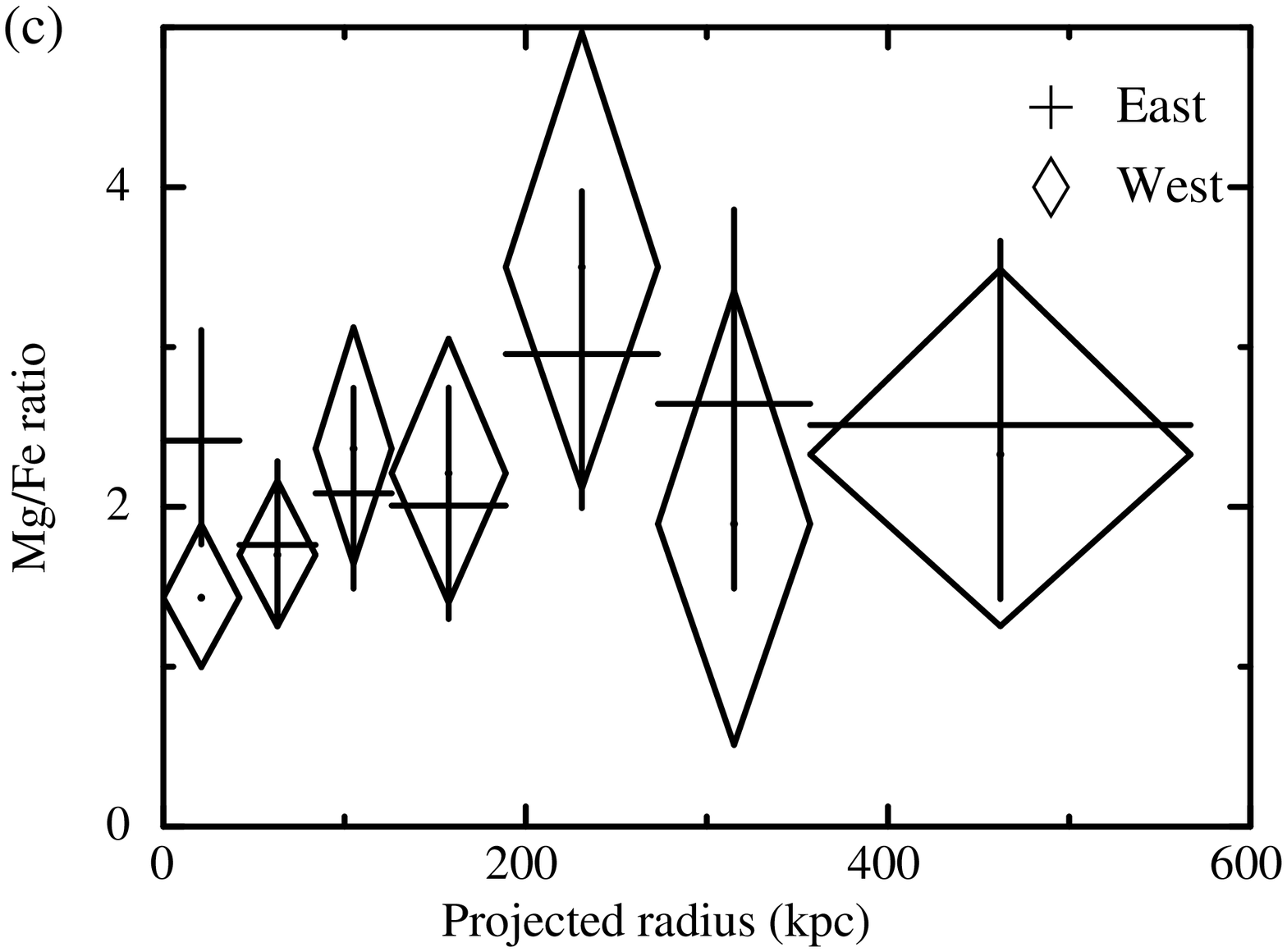}}
\end{minipage}
\hfill
\begin{minipage}{0.45\textwidth}
\centerline{\FigureFile(\textwidth,\textwidth){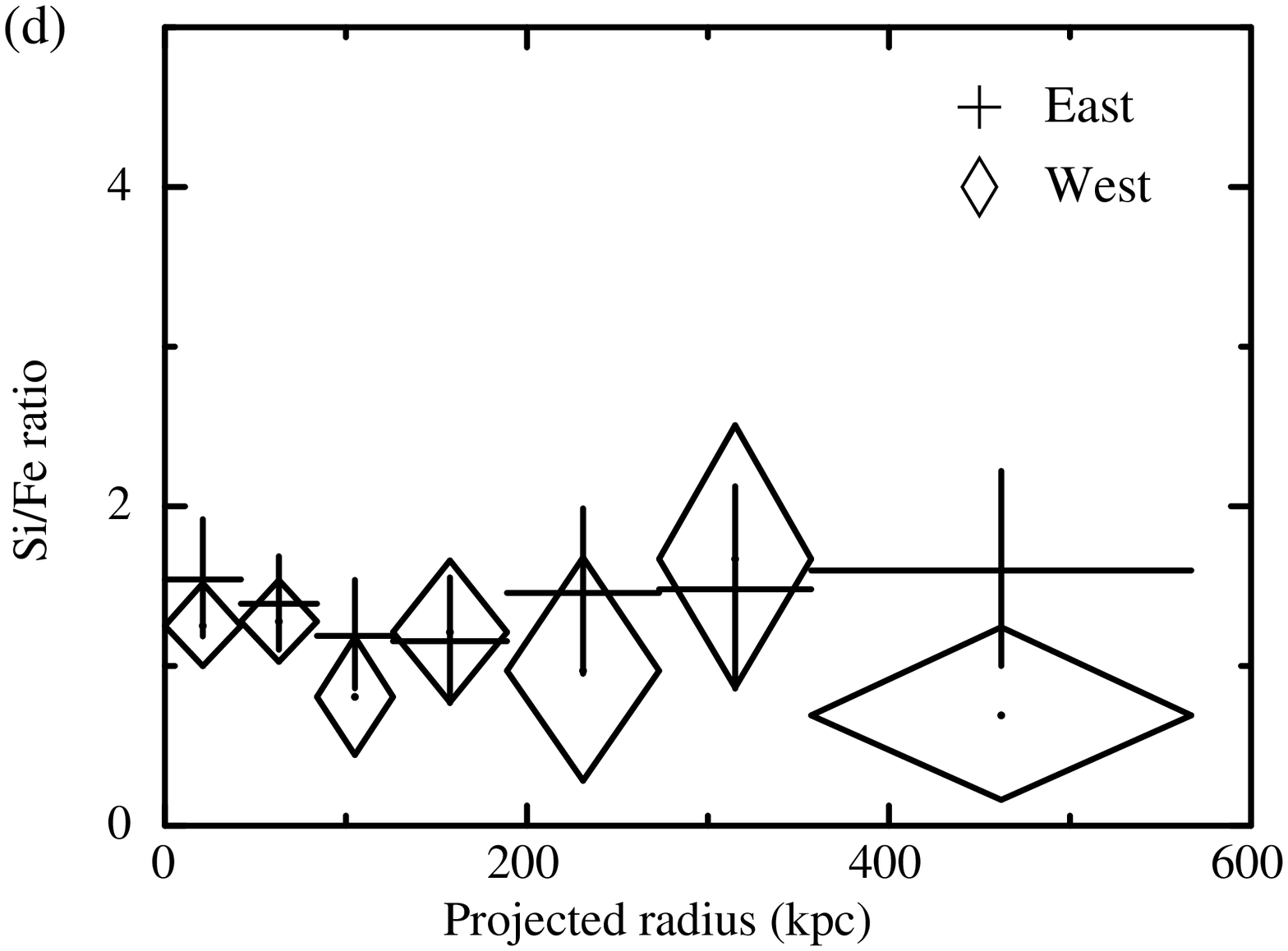}}
\end{minipage}
\end{center}
\caption{(a) Radial temperature profile along the east-west 
direction. (b)--(d) Radial O, Mg, and Si to Fe abundance ratios 
along with the east and west directions same as (a).
}\label{fig:5}
\end{figure*}

The fits are not acceptable due mainly to the very high photon
statistics compared with systematic errors in the instrumental
response.  However, these results are useful to assess how effectively
the metal abundances are constrained.  We fitted the spectra with all
the metal abundances as free parameters.  Somehow, Ne and Ni
abundances turn out to be higher than the other elements.  This might
be because the lines from these elements are not well resolved from
the Fe-L line complex.  We tentatively left these Ne and Ni abundances
vary freely in the spectral fit.

The resultant radial profiles of temperature and abundance are shown
in figure~\ref{fig:4}.  The temperature profile shows a mild cooling
core in the central region ($r\lesssim 6'$), and declines from $\sim
3.8$~keV at 6--9$'$ region ($\sim 130$~kpc) to $\sim 3.4$~keV in the
outermost annulus of 17--27$'$ ($\sim 360$--570~kpc $\sim
0.22$--0.35$\;r_{180}$).  Abundance profiles of O, Mg, Si, S, and Fe
are obtained up to a radius of $27'\simeq 570$~kpc.  Abundances of Mg,
Si, S, and Fe all decline with radius from $\sim 1.5$, 1.1, 1.1 and
0.9 solar in the central region, to $\sim 0.9$, 0.4, 0.3 and 0.4
solar, respectively, in the cluster outskirts.  The O abundance shows
somewhat flatter radial distribution.

Additional uncertainties on the derived parameters due to the OBF
contaminant and the background (CXB + NXB) systematics are shown in
table~\ref{tab:6} and figure~\ref{fig:4}.  Here, we fitted the spectra
again by changing the background normalization by $\pm 10$\%, and the
range of parameters are plotted with light-gray dotted lines in
figure~\ref{fig:4}. Systematic error in the background estimation is
almost negligible.  Change of the best-fit parameters by using
alternative modeling of the Galactic component with a single
temperature {\it apec} model was investigated, and the results are
indicated with light-gray dash-doted lines.  The differences are within the
statistical error for the inner five annuli. However, O abundance
becomes lower than the 90\% confidence error range at the outer two
annuli, and temperature and Fe abundance become lower at the outermost
annulus.  This is due mainly to the fact that the XIS cannot resolve
the ICM O\emissiontype{VIII} line from the Galactic one by the
redshift.  The systematic error range due to the uncertainty in the
OBF contaminant is indicated by black dotted lines.  A list of
$\chi^2$/dof is presented in table \ref{tab:7}.

\subsection{Difference Between East and West Regions}

We examined the difference in the temperature and abundance profiles
between the east and west directions from the cluster center.  We
treated the spectra in the east and west regions separately, but they
were always fitted jointly with the outermost spectrum in 17--27$'$
where the east and west data were combined together in the same way as
described in \ref{subsec:galactic}.  Figure \ref{fig:5}(a) shows the
temperature profiles along the east-west direction of the cluster.
The west direction shows slightly steeper temperature gradient than
the east direction.  Figure \ref{fig:5}(b)--(d) also show the
abundance ratios of O, Mg, Si divided by Fe in order to look into
relative variation in the abundance profiles.  While Si/Fe ratio is
consistent to be a constant value along the east-west direction, Mg/Fe
ratio looks an increase with radius in both the directions.  Though
O/Fe ratio has a large uncertainty, the west direction shows lager
values than the east.

\section{Bulk Motions in the ICM}

\begin{figure*}
\begin{center}
\begin{minipage}{0.48\textwidth}
\centerline{\FigureFile(\textwidth,\textwidth){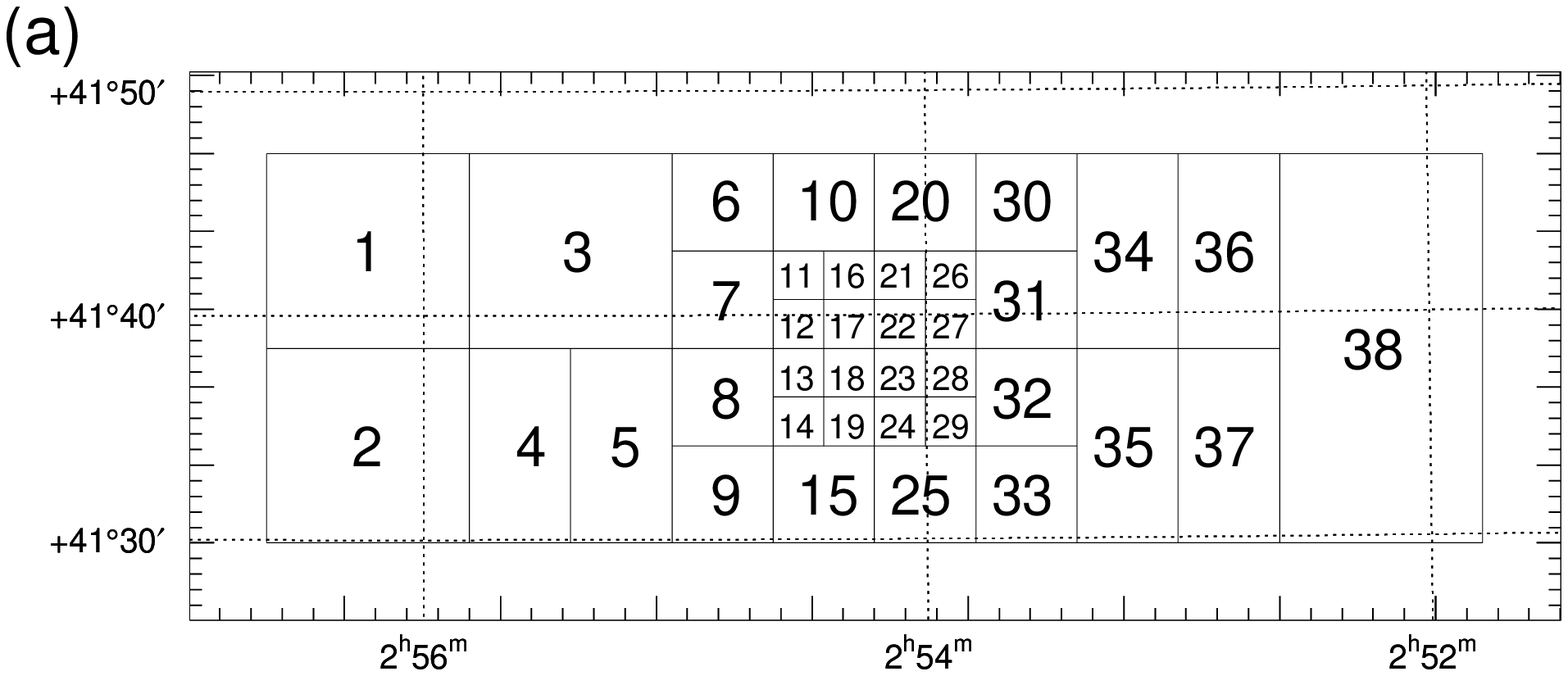}}
\end{minipage}
\hfill
\begin{minipage}{0.48\textwidth}
\centerline{\FigureFile(\textwidth,\textwidth){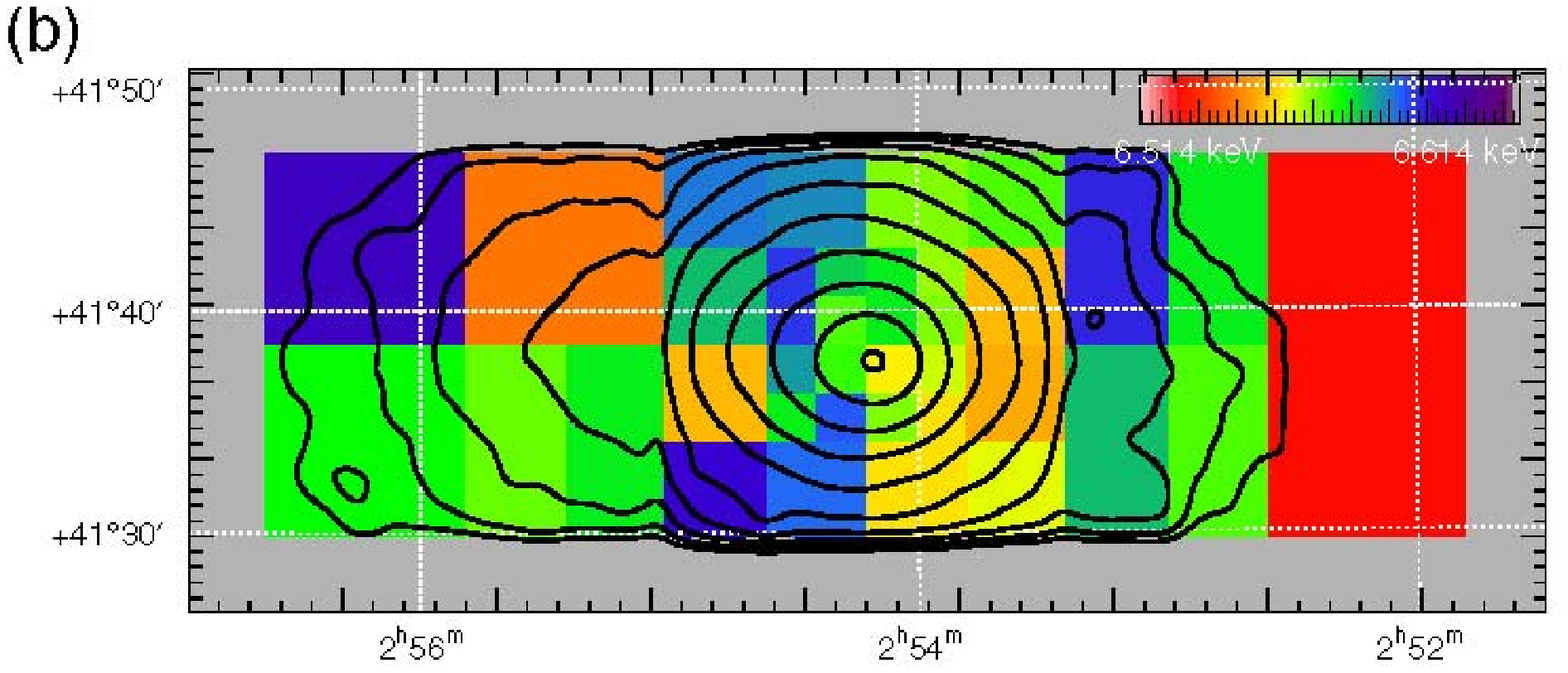}}
\end{minipage}
\begin{minipage}{0.80\textwidth}
\centerline{\FigureFile(\textwidth,\textwidth){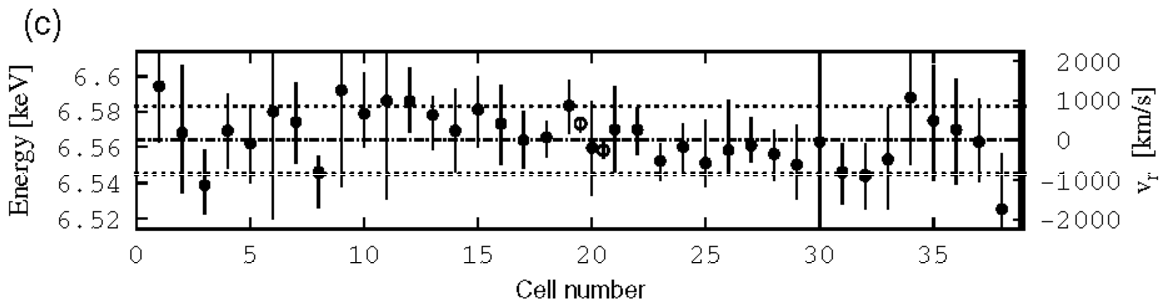}}
\end{minipage}
\caption{ Measurement of bulk motion using the central energy of
He-like Fe-K lines.  (a) Spatial division and the position
numbers. Depending of the brightness, the whole observed region is
divided into $2.2'\times2.2'$ (16 cells) or $4.5'\times4.5'$ (12
cells) regions for the central observation, and into $4.5'\times8.9'$
(6 cells), $8.9'\times8.9'$ (3 cells), and $17.8'\times8.9'$ (1 cell)
regions for the offset observations.  (b) Color representation of the
spatial distribution of the central energy of redshifted iron line for
each cell.  The map is overlaid on the contour in figure \ref{fig:1}.
(c) The energy of He-like Fe-K line against the cell numbers.  The
energy of redshifted He-like Fe-K$\alpha$ line at AWM~7 ($z=0.01724$)
is 6.566 keV\@. The dot-dashed line corresponds to the cluster
redshift, and the dotted lines indicate calibration error range
($z=0.003$ or $\sim20$~eV) of XIS\@.  Open circles indicate the
average for the east or west region in the central pointing.
}\label{fig:6}
\end{center}
\end{figure*}

\subsection{Spectral Analysis in Search of Bulk Motions}

In order to search for possible bulk motions of the ICM, the central
energy of He-like Fe-K line (the rest-frame energy of 6.679 keV) was
examined to look into a positional variation.  We divided the
$18'\times18'$ square XIS field of view into smaller cells. The
central observation was divided into 16 times $2.2'\times 2.2'$ cells,
12 times $4.5'\times 4.5'$ cells as shown in figure \ref{fig:6}(a).
For the east and west offset regions, we divided each field into 6
times $4.5'\times 8.9'$, 3 times $8.9'\times 8.9'$ and one
$17.8'\times 8.9'$ cells.  We co-added the data from only the FI
sensors (XIS0, 2 and 3).  For the accurate determination of the
iron-line energy, we fitted the background-subtracted XIS (FI) spectra
with a simple model over the 5--10 keV energy range. The model we used
consists of a continuum represented by a {\it power-law} model and
Gaussian profiles as described in \citet{ota07}, and we determined the
central energy of He-like Fe-K$\alpha$ line.  In the fit, we fixed the
intrinsic width of the Gaussian line to be 0.  We employed the central
energies of He-like Fe-K$\alpha$ and H-like Fe-K$\alpha$ lines at the
rest frame to be 6.679 and 6.964 keV, respectively, which are expected
from an {\it apec} model with $kT = 3.5$ keV\@.

\subsection{Constraint on the Existence of Bulk Motions}
\label{subsec:bulk}

Distribution of the line central energy are shown as a color coded map
in figure \ref{fig:6}(b) and against the cell number in 
figure \ref{fig:6}(c), with 90\%
confidence statistical errors.  The dot-dashed line in figure
\ref{fig:6}(c) corresponds to the redshifted central energy at AWM~7
of $E_{\rm cl}= 6.566$ keV with $z=0.01724$, based on the rest-frame
central energy $E_0=6.679$ keV\@.  The dotted lines show the range of
current calibration error of $z=0.003$ or $\sim 20$ eV\@.  Figure
\ref{fig:6}(b) and (c) indicate no significant systematic feature for
the gas motion within the 90\% confidence level.

Since there seem to
be some hint of difference between the east and west regions in the
central observation, we looked into a larger structure.
Spectra in the cell numbers of 6--19, or 20--33
were summed up to evaluate the east or west difference
for the central pointing, respectively.
The results for the east and west regions correspond to
the left and right open circles in figure~\ref{fig:6}(c).
The difference was $15\pm 6$~eV between the east and west
regions with the 90\% confidence error.
However, it is known that the central energy of a line shows
dependence on the read-out direction of XIS chip,
and the systematic uncertainty is about $\sim 20$~eV\@.
We therefore conclude that the upper limit for the bulk motion of
the gas to be $\Delta E\lesssim 40$ ($15\pm 6\pm 20$) eV
based on the measurement of the central energy of He-like Fe-K$\alpha$ line,
which corresponds to $\Delta v \lesssim 2000$ km~s$^{-1}$.

\section{Discussion}\label{sec:discuss}

\begin{figure}
\begin{minipage}{0.45\textwidth}
\centerline{
\FigureFile(\textwidth,\textwidth){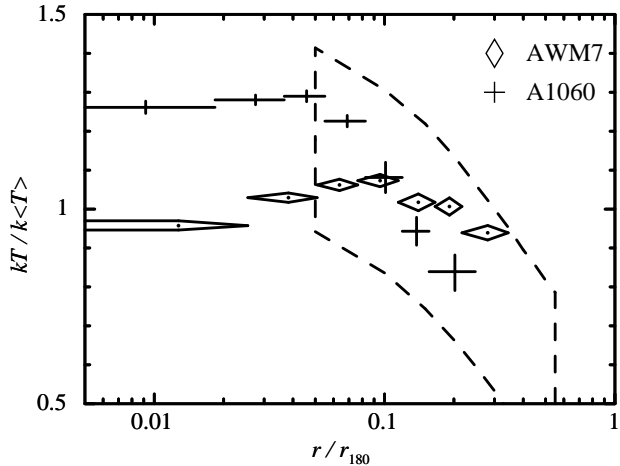}}
\caption{ Scaled temperature profiles using normalized temperatures
and the radius scaled by the virial radius.  The A~1060 data are taken
from \citet{sato07}.  The region indicated by dashed line shows 
result of ASCA observations by \citet{markevitch98}.  }\label{fig:7}
\end{minipage}
\end{figure}

\subsection{Temperature Profile}

Suzaku observation showed the temperature profile of AWM~7 to 
$\sim 0.3 r_{\rm 180}$ region of the cluster, extending the previous
Chandra and XMM-Newton measurements.  \citet{hayakawa06} showed the
temperature in AWM~7 to drop from 3.8 keV at the peak to 3.5 keV at
$r\sim13'$ with XMM-Newton, and the present study confirmed their
results in a wider range of radius.  Previous ASCA observation of AWM~7
indicated fairly uniform temperature distribution in the whole cluster
region \citep{furusho01,ezawa97}, however their results are consistent
with ours within the error because the temperature drop is quite small
that it could well be regarded as isothermal in observation with lower
statistics.

To compare the temperature profile with those in other clusters,
we normalized temperature with the emission-weighted average
of $k\langle T\rangle = 3.56$~keV for AWM7,
and plotted it against the radius normalized with the viral radius
in figure \ref{fig:7}.
We calculated the $k\langle T\rangle$ in the range of
0.1--$0.3\; r_{180}$ with Suzaku.
The normalized temperature profile is compared with Abell~1060,
which shows $k\langle T\rangle = 2.49$~keV in 0.1--$0.25\; r_{180}$
\citep{sato07}.
Note that AWM~7 has a cD galaxy at the center,
while Abell~1060 does not. The temperature gradient of AWM~7 is
clearly less steep than that of Abell~1060, while these profiles are
both consistent with the broad range (dashed line region in figure
\ref{fig:7}) given by \citet{markevitch98} with ASCA\@.  The
\citet{markevitch98} result is consistent with the temperature
gradients for several clusters studied by \citet{vikhlinin05} with
Chandra and \citet{piffaretti05} with XMM-Newton.

\citet{hayakawa06} showed a systematic difference with XMM-Newton in
the temperature gradients between 9 cD and 13 non-cD clusters,
including both AWM~7 and Abell~1060.  They showed that both cD and
non-cD clusters show consistent gradients for a radius range
$0.05<r/r_{180}<0.55$, with general agreement with the
\citet{markevitch98} result with slightly higher absolute
temperatures.  They, however, found that, in $r/r_{180}<0.05$,
temperatures of the cD clusters showed a systematic drop toward the
center, and this feature was not clearly recognized in the non-cD
clusters.  Such a feature in the central region is also seen in our
observation of AWM~7.  

\subsection{Abundance Profiles}

The good XIS sensitivity to emission lines, especially below 1 keV,
enabled us to measure O and Mg abundances out to $\sim0.35~r_{180}$
with Suzaku.  The radial abundance profiles of Si, S, and Fe show
steeper drop than those of O and Mg.  \citet{ezawa97} reported a
large-scale abundance gradient from 0.5 solar at the center to
$\lesssim0.2$ solar in the region over 500 kpc ($H_0=50$ km
s$^{-1}$ Mpc$^{-1}$), consistent with our result for Fe.  On the other
hand, abundance determination of Ni and Ne has a problem because of
possible confusion with strong Fe-L lines.  Also, we have to note that
the O abundance in the outer region of our observation is strongly 
affected by the foreground Galactic emission as described in subsection
\ref{subsec:galactic}. Even allowing the O distribution
to be a rather preliminary result, the combined Mg and O
feature suggests that their distribution is likely to be more extended
than those of Fe and Si. This is at least consistent with the view
that the metals injected from Type II supernova, i.e.\ O and Mg, in
the early phase of galaxy formation in the wind form show extended
distribution \citep{arimoto87}.
In contrast, the steep gradient of Fe and Si distribution suggests they
are mostly produced by Type Ia supernova in a later stage, with
enhanced contribution from the cD galaxy as shown by
\citet{degrandi04}.

We calculated cumulative metal mass profiles by combining the present
abundance results with X-ray gas mass profile with XMM-Newton.  Since
measured metals have different yields from the two types of supernova,
with O and Mg mostly synthesized by type II supernovae (SNe II) while
type Ia supernovae (SNe Ia) giving large yields of Si, S and Fe, we
can estimate their relative contributions which can best account for
the measured abundance pattern.  We thus estimated the number ratio of
SNe II to SNe Ia to be $\sim4$ in the ICM of AWM~7 as reported in
\citet{sato07b}.  We also calculated radial mass-to-light ratios for
O, Mg, and Fe, and compared the values with those in other systems.
As a result, it is suggested that smaller systems with lower gas
temperature tend to show lower mass-to-light ratios for O, Mg, and Fe
as shown in \citet{sato07c}.

\subsection{Morphology of AWM~7}

No significant bulk motion of the gas exceeding the sound speed,
$\sim1000$ km s$^{-1}$, was detected in the ICM of AWM~7, even though
the X-ray image of AWM~7 is highly elongated in the east-west
direction.
\citet{neumann95} and \citet{furusho01} reported the major to minor
axis ratio to be 0.8. This elongation is fairly parallel to the
large-scale filament of the Pisces-Perseus supercluster, which runs
almost along the east-west direction.  We will consider possible
processes to explain this elongation.  In general, the simplest
explanation for the deviation from the spherical symmetric shape is
that the cluster is rotating. However, X-ray spectra show no
indication of significant rotation at present. Furthermore, based on
the standard CDM scenario, it is difficult to make a cluster to have a
large angular momentum \citep[e.g.][]{peebles69,white84,barnes87}.
Tidal interaction with the Perseus cluster is another possible process, 
however \citet{neumann95} showed that the tidal effect of Perseus cluster was
too small to cause the observed large ellipticity of AWM~7.

Another possible process is asymmetric accretion of matter \citep[see
e.g.][]{neumann95}. Directional mass accretion may lead to an
ellipsoidal or triaxial dark matter halo. If the intra-cluster gas is
in hydrostatic equilibrium in the gravitational potential dominated by
the dark matter, gas distribution should deviate from the spherical
symmetry.  If we can obtain more detailed information about the
profile of the eccentricity of the gas isodensity surface as well as
precise density and temperature profiles, we can reconstruct the
three-dimensional dark matter structure \citep[e.g.][]{lee04}. This
will give us an important knowledge of the dark matter relaxation
process.

Finally, if the gas infalls in an asymmetric manner from the filament
onto the relaxed cluster and the accreted gas has not been mixed with
the original intra-cluster gas, the gas profile deviates from
spherical symmetry and X-ray morphology becomes elliptical.  
In this case, the gas in the outer part of the east and west regions may keep
the physical properties such as metal abundance when the gas was in
the filament. 
The warm gas in the
large-scale filament is an important part of the warm-hot
intergalactic medium (WHIM) which is the dominant component of baryons
in the local universe \citep[e.g.][]{yoshikawa03}. Several missions
are proposed to detect WHIM through O VII and O VIII emission lines
\citep[e.g.][]{ishisaki04,ohashi06,piro06,denherder06}.  
If metal abundance of gas in
filaments is as high as that of ICM, the detection of WHIM becomes
relatively easy and the missing baryon problem can be answered in the
near future. Furthermore, we need to revisit the metal production and
ejection processes closely to explain the high metal abundance in the
intergalactic gas. To confirm such a possibility, deeper X-ray
observation in the outermost region of AWM~7 is important.

\bigskip
Part of this work was financially supported by the Ministry of
Education, Culture, Sports, Science and Technology of Japan,
Grant-in-Aid for Scientific Research
No.\ 14079103, 15340088, 15001002, 16340077, 18740011.

\end{document}